\documentclass[aps,prb,twocolumn,showpacs,final,floatfix]{revtex4-1}
\usepackage{nicematrix}
\usepackage{amsmath}
\usepackage{amsfonts}
\usepackage{amssymb}
\usepackage{physics}
\usepackage{color}
\usepackage{hyperref}
\usepackage{inputenc}
\usepackage{enumerate}
\usepackage{graphicx}
\usepackage{float}
\usepackage{amsmath}
\usepackage{amsfonts}
\usepackage{setspace} 
\usepackage{lipsum}
\usepackage{dcolumn}
\usepackage{hyperref}
\usepackage{subfigure}
\usepackage{epsfig}
\usepackage{epsf}
\usepackage{epstopdf}

\usepackage{color}
\usepackage{enumitem}
\usepackage{bm}
\usepackage{mathrsfs}
\begin{document}

% Use the \preprint command to place your local institutional report
% number in the upper righthand corner of the title page in preprint mode.
% Multiple \preprint commands are allowed.
% Use the 'preprintnumbers' class option to override journal defaults
% to display numbers if necessary
%\preprint{}

%Title of paper
\title{Equivalence of  NEGF  and scattering approaches to electron transport in the Kitaev chain }

% repeat the \author .. \affiliation  etc. as needed
% \email, \thanks, \homepage, \altaffiliation all apply to the current
% author. Explanatory text should go in the []'s, actual e-mail
% address or url should go in the {}'s for \email and \homepage.
% Please use the appropriate macro foreach each type of information
% \affiliation command applies to all authors since the last
% \affiliation command. The \affiliation command should follow the
% other information
% \affiliation can be followed by \email, \homepage, \thanks as well.
%\email[]{Your e-mail address}
%\homepage[]{Your web page}
%\thanks{}
%\altaffiliation{}
\author{Junaid Majeed Bhat}
\author{Abhishek Dhar}

\affiliation{International Centre for Theoretical Sciences,  Tata Institute of Fundamental Research, Bengaluru-560089, India}
%Collaboration name if desired We present a stochastic formulation of the Kadanoff-Baym or Keldysh theory to calculate the conductance

%option in \documentclass). \noaffiliation is required (may also be
%used with the \author command).
%\collaboration can be followed by \email, \homepage, \thanks as well.
%\collaboration{}
%\noaffiliation

\date{\today}

\begin{abstract}
 We consider electron transport in a Kitaev chain connected at its two ends to normal metallic leads kept at different temperatures and chemical potentials. Transport in this set-up is usually studied using two frameworks ---  the nonequilibrium Green's function (NEGF) approach or the scattering approach. In the NEGF approach the current and other steady state properties of a system are expressed in terms of Green's functions that involve the wire properties and self-energy corrections arising from the leads. In the scattering approach,     transport is studied in terms of the scattering amplitudes of  plane waves incident on the wire from the reservoirs. 
Here we show explicitly  that these two approaches produce  identical results for the conductance of the Kitaev chain. Further we show that 
the NEGF expression for conductance can be written in such a way that there is a one-to-one correspondence of the various terms in the NEGF expression to the  amplitudes for normal transmission, Andreev transmission and Andreev reflection in the scattering approach. Thereby, we obtain closed form expressions for these. We  obtain the wavefunctions of zero energy Majorana bound states(MBS) of the wire connected to leads and prove that they are present in the same parameter regime in which they 
occur for an isolated  wire. These bound states give rise to perfect Andreev reflection responsible for zero bias quantized conductance peak. We  discuss the dependence of the width of this  peak on different parameters of the Hamiltonian and relate it to the MBS wavefunction properties. We find that the peak broadens if the weight of the MBS in the reservoirs increases and vice versa. 
\end{abstract}

% insert suggested PACS numbers in braces on next line
% insert suggested keywords - APS authors don't need to do this
%\keywords{}
\maketitle
%\maketitle must follow title, authors, abstract, \pacs, and \keywords

% body of paper here - Use proper section commands
% References should be done using the \cite, \ref, and \label commands
\section{Introduction}
Electron transport properties of the Kitaev chain, a simple example of a  one-dimensional spinless superconducting wire, has  been extensively investigated recently~\cite{kitaev,sau2010non,oreg2010helical,mourik2012,das2012zero,diptiman2015}. Amongst the  interesting experimental results are the signatures of the so-called Majorana bound states (MBS) seen in measurements of the electrical conductance. Theoretically, transport has been studied in this system using the quantum Langevin equations - nonequilibrium Green's function approach~\cite{roy2012, roy2019,bhat2020transport} (QLE-NEGF), scattering approach~\cite{btk,diptiman2015,maiellaro2019} and the  Keldysh nonequilibrium  Green's function approach~\cite{lobos2015,doornenbal2015conductance,komnik2016,zhang2020}.  The  QLE-NEGF and Keldysh approach start from the same microscopic model of system-bath Hamiltonian with the Kitaev wire sandwiched between two normal leads (the baths), and involve elimination of bath degrees of freedom to find the steady state properties of the system. For normal (without superconducting pairing potential terms) wires, the equivlanece of these two approaches has been established quite generally~\cite{dhar2006}.  In both these approaches the conductance is given in terms of nonequilibrium Green's functions. On the other hand, in the scattering approach one considers scattering of  plane waves, incident from the normal metallic reservoirs, by the superconducting region. The conductance is expressed in terms of the scattering amplitudes.  For the case of the Kitaev chain and more generally in superconducting wires, three scattering processes are identified corresponding to normal transmission, Andreev transmission and Andreev reflection~\cite{btk,diptiman2015,maiellaro2019}. 
It is expected that the scattering formalism and the NEGF formalism should be equivalent and one of the main aims of the present paper is an explicit demonstration of this equivalence. 

 In the QLE-NEGF approach one writes the quantum Langevin equations of motion for the system and then the transport properties are found via its steady state solution. The quantum Langevin equation follows from removing the bath degrees of freedom from the Heisenberg equation of motion for the wire. The QLE-NEGF approach was first applied to the Kitaev model in Ref.~\onlinecite{roy2012}, while Ref.~\onlinecite{bhat2020transport}  provides a more general and complete application of this method to  spinless superconducting wires connected to two normal baths, obtaining explicit expressions for particle and energy currents. One of the earliest application of the scattering approach to  a system consisting of a metallic lead connected to a superconductor was by Blonder, Tinkham and Klapwijk~\cite{btk}.  More recent work has implemented the scattering approach to a 1-D superconductor sandwiched between two metallic leads~\cite{diptiman2015}. However, both of these papers consider the continuum models for the superconductor.  The scattering approach  has been used earlier to study superconducting lattice models~\cite{nehra2020},  but to our knowledge, its connection to the NEGF approach for the same model has not been explored so far. It has been understood that the NEGF particle current and  conductance at the ends of the superconductor has one local and two non-local contributions which are attributed to Andreev and normal scattering amplitudes~\cite{lobos2015,bhat2020transport}.  In this paper, we present  a detailed study of the scattering approach to the Kitaev chain which is a simple example of a 1-D spinless superconductor and  show analytically  that it yields the same results as the QLE-NEGF approach. We also show that the three terms in the NEGF current expression correspond precisely to different scattering processes that  take place in the system. The main idea is to express the Green's function in terms of transfer matrices from which the connection to the scattering process becomes clear. This treatment follows Ref.~\onlinecite{das2012landauer}  but now involves  $4\cross4$ transfer matrices instead of $2\cross2$ ones for normal 1-D wires.

An interesting aspect of the Kitaev chain is that, in the topologically non-trivial parameter regime they host special zero modes  called Majorana Bound states(MBS). These are symmetry protected robust states localized at the two edges of the wire. These states are responsible for a perfect Andreev reflection at zero bias and lead to the  zero bias peak in the conductance that has been observed experimentally~\cite{mourik2012,das2012evidence,aguado2017majorana}. This peak is is believed to be one of the strongest experimental signatures of the MBS. These states  for isolated  wires (in the absence of reservoirs) were first discussed by Kitaev~\cite{kitaev}. In the present work we explore the effect of the reservoirs on these exotic states  and  relate the properties of the MBS wavefuction, which now leaks into the leads,  to the  behaviour of the zero bias conductance peak width. We find that these states are present  in the same parameter regime in which they occur in the isolated wire and that the conductance peak broadens as the weight of the MBS wavefunction increases in the two reservoirs. 

This paper is structured as follows: in Sec.~\ref{model}, we introduce the exact lattice Hamiltonian  considered for the calculations and  provide a summary of the results from the QLE-NEGF and scattering approaches applied to this model. We also discuss qualitatively the expected  equivalence between the two approaches. In Sec.~\ref{scattering} we provide explicit details of the calculation involved in  scattering approach and also discuss the zero mode MBS in this system.  The analytical proof for the equivalence of the two approaches is given in  Sec.~\ref{sec_proof}. A  numerical demonstration of this is provided in Sec.~\ref{numerical} along with a discussion of the behaviour of the conductance peak and its relation to the MBS wavefunction. We conclude in Sec.~\ref{sec_concl}.     

\section{The Model and the equivalence of the two Approaches}
\label{model}
 In this section, we introduce the model for the Kiteav chain connected to reservoirs at its two ends and then summarize the results obtained by applying the QLE-NEGF approach and the scattering approach to this model. After that, we qualitatively discuss the equivalence of the two approaches which is later proven analytically in Sec.~\ref{sec_proof}.  The  Hamiltonian of the Kitaev chain(1-D wire), $\mathcal{H}_W$, is given by normal  tight binding Hamiltonian  with  mean field BCS-type coupling between its neighbouring sites. The reservoirs are taken to be semi-infinite chains with nearest neighbour tight binding Hamiltonians, $\mathcal{H}_L$ and $\mathcal{H}_R$. $L$ and $R$ refer to the left and the right reservoir respectively. The finite ends of the reservoirs are placed at ends of the wire and its extremal sites are coupled to the nearest reservoir sites via tight binding Hamiltonians, $\mathcal{H}_{WL}$ and $\mathcal{H}_{WR}$. The creation and annihilation operators, satisfying usual fermionic anti-commutation relations, for the wire, the left bath and the right bath are denoted as $\lbrace c_j^\dagger, c_j\rbrace$, $\lbrace c_\alpha^\dagger,c_\alpha\rbrace$ and $\lbrace c_{\alpha'}^\dagger,c_{\alpha'}\rbrace$ respectively. The Latin indices $j,k,..$ are taken to label the sites on the wire. These take values from $1,2,...,N$, $N$ being the number of sites on the wire. Similarly,  Greek indices $\alpha, \nu,..$ taking values from $-\infty,...,-1,0$ and primed Greek indices $\alpha',\nu',..$ taking values from $N+1,N+2,...,\infty$ label the left reservoir and right reservoir sites respectively. The full Hamiltonian is thus given by,
 \begin{align}
 	\mathcal{H}&=\mathcal{H}_W+\mathcal{H}_{WL}+\mathcal{H}_{WR}+\mathcal{H}_{L}+\mathcal{H}_{R}\label{H},\\
 \text{where}~~ \mathcal{H}_W&=\sum_{j=1}^{N-1} \left[-\mu_w c_j^\dagger c_j-\eta_w(c_j^\dagger c_{j+1}+c_{j+1}^\dagger c_j)\right.\notag\\&\left.\hspace{2cm}+\Delta(c_jc_{j+1}+c_{j+1}^\dagger c_j^\dagger )\right],\label{H_W}\\
 \mathcal{H}_{WL}&=-\eta_c(c_{1}^\dagger c_{0}+c_0^{\dagger} c_{1}),\label{H_WL}\\
 \mathcal{H}_{WR}&=-\eta_c(c_{N}^\dagger c_{N+1}+c_{N+1}^{\dagger} c_{N}),\label{H_WR}\\
 \mathcal{H}_L&=-\eta_b\sum_{\alpha=-\infty}^{0}c_\alpha^{\dagger} c_{\alpha+1}+c_{\alpha+1}^{\dagger} c_\alpha\label{H_L},\\
 \mathcal{H}_R&=-\eta_b\sum_{\alpha'=N+1}^{\infty}c_{\alpha'}^{\dagger} c_{\alpha'+1}+c_{\alpha'+1}^{\dagger} c_{\alpha'}\label{H_R},
\end{align}
where $\Delta,~\eta_w,~\mu_w$ are respectively the superconducting pairing strength, hopping amplitude and the chemical potential on the sites of the wire, $\eta_c$ is the coupling strength between the wire and the reservoirs, and the hopping amplitude in the reservoirs is given by $\eta_b$.  For simplicity, all of these parameters are taken to be real. The reservoirs are initially described by grand canonical ensembles at temperatures, $T_L, T_R$ and chemical potentials, $\mu_L,\mu_R$ and,  as we will see, this determines the correlation properties of the noise terms in the final Langevin equations.

We first present the QLE-NEGF results for electron transport in this model. Following the steps given in Ref.~\onlinecite{dhar2006,bhat2020transport}, we  start from the Heisenberg equations of motion for the entire system which given by,
\begin{align}
\notag\dot{c}_l&=-i\sum_m H^W_{lm}c_m-i\sum_mK_{lm}c_m^\dagger\\&\hspace{2cm}-i\sum_{\alpha}V^L_{l\alpha }c_\alpha-i\sum_{\alpha^\prime}V^R_{l\alpha^\prime }c_{\alpha^\prime},\label{cldot}\\
\dot{c}_{\alpha}&=-i\sum_{\nu}H^L_{\alpha\nu}c_{\nu}-i\sum_l V^{L\dagger}_{\alpha l}c_l\label{calphadot},\\
\dot{c}_{\alpha^\prime}&=-i\sum_{\nu^\prime}H^R_{\alpha^\prime\nu^\prime}c_{\nu}-i\sum_l V^{R\dagger}_{\alpha^\prime l}c_l,\label{calpha'dot}
\end{align}
where        
\begin{align}
H^W_{lm}&=-\mu_w\delta_{lm}-\eta_w(\delta_{l,m-1}+\delta_{l,m+1}),\label{Hwlm}\\
K_{lm}&=\Delta(\delta_{l,m+1}-\delta_{l,m-1}),\label{Klm}
\end{align}
\begin{align}
H^L_{\alpha\nu}&=-\eta_b(\delta_{\alpha,\nu-1}+\delta_{\alpha,\nu+1}),\label{HL}\\
H^R_{\alpha'\nu'}&=-\eta_b(\delta_{\alpha',\nu'-1}+\delta_{\alpha',\nu'+1}),\label{HR}\\
V^L_{l\alpha}= -\eta_c &\delta_{l1}\delta_{\alpha0}~~~V^R_{l\alpha'}= -\eta_c \delta_{lN}\delta_{\alpha',N+1}.\label{VLVR}
\end{align}
The time dependent bath degrees of freedom can be  removed from the Heisenberg equation for wire operators using appropriate Green's functions, namely
\begin{align}
&g_L^+(t)=-ie^{-itH^L}\theta(t)=\int_{-\infty}^\infty \frac{d\omega}{2\pi}g_L^+(\omega)e^{-i\omega t}\label{gL}\\
\text{and}~&g_R^+(t)=-ie^{-itH^R}\theta(t)=\int_{-\infty}^\infty \frac{d\omega}{2\pi}g_R^+(\omega)e^{-i\omega t}\label{gR},
\end{align}
for the left and right reservoirs respectively.  These two equations furnish formal solutions for the Hiesenberg equations of motion for the reservoir operators. These solutions can be used to eliminate the reservoir degrees of freedom from the Heisenberg equation for the wire operators to give the quantum Langevin equation for the wire~\cite{bhat2020transport}:
\begin{align}
&\label{les}i\dot{c}_l=\sum_m H^W_{lm}c_m+\sum_mK_{lm}c_m^\dagger+\eta^L_l(t)+\eta_l^R(t) \nonumber \\&+\int_{-\infty}^{t}ds\left([\Sigma^+_L(t-s)]_{l m} +[\Sigma^+_{R}(t-s)]_{l m}\right) c_m(s).
\end{align}
Thus, the effect of the reservoirs on the dynamics of the wire operators is expressed as the sum of the noise $\eta^L_l(t)$ and $\eta^R_l(t)$
and the history dependent dissipation terms given by the integrals. Here $\Sigma^+_{L}(t)=V^Lg^+_L(t)V^{L\dagger}$ and $\Sigma^+_R(t)=V^Rg^+_R(t)V^{R\dagger}$ are therefore the self energy corrections to the wire due to the left and the right reservoirs respectively.  The properties of the noise and dissipation are easiest to express in Fourier space and are given by:
\begin{align}
&\Sigma_{L}^+(\omega)=V^{L}g_L^+(\omega)V^{L\dagger}, \label{SigmaLR} \\
&\expval{\tilde{\eta}_l^{L\dagger}(\omega)\tilde{\eta}_m^L(\omega^\prime)}=[\Gamma_L(\omega)]_{ml}f_L(\omega)\delta(\omega-\omega^\prime)\label{bcorr1},\\
&\expval{\tilde{\eta}_l^{L}(\omega)\tilde{\eta}_m^{L\dagger}(\omega^\prime)}=[\Gamma_L(\omega)]_{lm}\left[1-f_L(\omega)\right]\delta(\omega-\omega^\prime).\label{bcorr2}
\end{align}
with $\Gamma_{L}=\frac{1}{2\pi i}(\Sigma_{L}^-(\omega)-\Sigma^+_{L}(\omega))$ and $f_L(\omega)=f(\omega,\mu_L,T_L)=[e^{(\omega-\mu_L)/T_L}+1]^{-1}$ is the usual Fermi-Dirac distribution. The right reservoir will have similar properties.  

Our goal is now to obtain the steady state solution for the wire operators  which would then give us the steady state current entering the wire from the left reservoir. To that end, we assume a parameter regime of the Hamiltonian which allows a steady state~\cite{dhar2006, bhat2020transport} and  the corresponding solution can be written by taking a  Fourier transform of Eq.~(\ref{les}) to give
\begin{equation}
[\Pi(\omega)]_{lm}\tilde{c}_m(\omega)-K_{lm}\tilde{c}_m^\dagger(-\omega)=\tilde{\eta}_l^L(\omega)+\tilde{\eta}_l^R(\omega), \label{fts}
\end{equation}
with $\Pi(\omega)=\omega-H^W-\Sigma^+_L(\omega)-\Sigma^+_R(\omega).\label{Pi}$
 We note that Eq.~(\ref{fts}) forms a set of linear equations in variables $c_l(\omega)$ and $c_l^\dagger(-\omega)$ and therefore we can solve for  $c_l(\omega)$ to get
\begin{align}
\notag\tilde{c}_m(\omega)&=[G_1^+(\omega)]_{ml} \left[\tilde{\eta}_l^L(\omega)+\tilde{\eta}_l^R(\omega)\right]\\&+[G_2^+(\omega)]_{ml} \left[\tilde{\eta}_l^{L\dagger}(-\omega)+\tilde{\eta}_l^{R\dagger}(-\omega)\right],\label{ngef}
\end{align}
where $G_1^+(\omega)$ and $G_2^+(\omega)$ are defined as
\begin{align}
G_1^+(\omega)&=\frac{1}{\Pi(\omega)+K[\Pi^*(-\omega)]^{-1}K^\dagger},\label{G1}\\
G_2^+(\omega)&=G_1^+(\omega)K[\Pi^*(-\omega)]^{-1}.\label{G2}
\end{align}
The steady state solution is now expressed in terms of the two Green's functions $G_1^+(\omega)$ and $G_2^+(\omega)$.   The steady state properties of the wire would be given in terms of these two Green's functions and would involve the correlation properties of the noise terms which are determined by the initial states of the reservoirs.

 The  current coming from the left reservoir, $J_L$ can be obtained from  the rate of change of total number of particles in the left reservoir and is given by,
 \begin{equation}
 J_L=2\sum_{m\alpha}\Im[V_{m\alpha}^L\expval{c_m^\dagger(t) c_{\alpha}(t)}]\label{currfr}
 \end{equation}
 Using Eq.~(\ref{ngef}) and the correlation properties of the reservoirs, Eqs.~\ref{bcorr1},\ref{bcorr2}), one  finds~\cite{bhat2020transport} that Eq.~(\ref{currfr}) leads to the following expression for the current:
 \begin{align}
 \notag &J_L=\int_{-\infty}^{\infty}\frac{d\omega}{2\pi} \bigg(T_1(\omega)(f_L^e(\omega)-f_R^e(\omega))\\&+T_2(\omega)(f_L^e(\omega)-f_R^h(\omega))+T_3(\omega)(f_L^e(\omega)-f_L^h(\omega))\bigg),\label{currexp1}
 \end{align}
where $f_X^e(\omega)=f(\omega,\mu_X,T_X)$, $f_X^h(\omega)=f(\omega,-\mu_X,T_X)$, $(X=L,R)$ are the electron and hole occupation numbers and
\begin{align}
&T_1(\omega)=4\pi^2\Tr[G_1^+(\omega)\Gamma_R(\omega)G_1^{-}(\omega)\Gamma_L(\omega)],\\
&T_2(\omega)=4\pi^2\Tr[G_2^+(\omega)\Gamma_R^T(-\omega)G_2^{-}(\omega)\Gamma_L(\omega)],\\
&T_3(\omega)=4\pi^2\Tr[G_2^+(\omega)\Gamma_L^T(-\omega)G_2^{-}(\omega)\Gamma_L(\omega)].
\end{align}
From the expression for current we can obtain the conductance at the left end and, in  units of $e^2/h$, is found to be:
\begin{equation}
G_L(T_L,\mu_L)=2\pi \pdv{J_L}{\mu_L},
\end{equation}
which at, $T_L=T_R=0$, gives:
\begin{equation}
G_L=T_1(\mu_L)+T_2(\mu_L)+T_3(\mu_L)+T_3(-\mu_L). \label{GLnegf1}
\end{equation}
The transmission functions involve the two Green's functions $G_1^+(\omega)$ and $G_2^+(\omega)$ given by Eq.~(\ref{G1}) and Eq.~(\ref{G2}) respectively. The various matrices which are present in their expression have simple forms and can be obtained explicitly by using Eq.~(\ref{Hwlm}-\ref{VLVR}):
\begin{align}
[\Sigma^+_{L}(\omega)]_{ij}&=\eta_c^2 \Sigma(\omega) \delta_{i1}\delta_{j1},\\
[\Gamma_{L}(\omega)]_{ij}&=\frac{\eta_c^2}{\pi} g(\omega) \delta_{i1}\delta_{j1},  \\ 
[\Sigma^+_{R}(\omega)]_{ij}&=\eta_c^2 \Sigma(\omega)\delta_{iN}\delta_{jN},\\
[\Gamma_{R}(\omega)]_{ij}&=\frac{\eta_c^2}{\pi} g(\omega)\delta_{iN}\delta_{jN},\\
[\Pi(\omega)]_{ij}&=(\omega+\mu_w)\delta_{ij}+\eta_w(\delta_{i,j+1}+\delta_{i,j-1})\notag\\& -\eta_c^2 \Sigma(\omega)\delta_{i1}\delta_{j1}-\eta_c^2 \Sigma(\omega)\delta_{iN}\delta_{jN}\label{Piij},
\end{align} 
where $g(\omega)=\Im[\Sigma(\omega)]$, and it can be shown that~\cite{dhar2006}
\begin{align}
\Sigma(\omega)= 
\begin{cases}
\frac{1}{\eta_b}\left(\frac{\omega}{2\eta_b}-\sqrt{\frac{\omega^2}{4\eta_b^2}-1}\right),& \text{if } \omega > 2\eta_b\\
\frac{1}{\eta_b}\left(\frac{\omega}{2\eta_b}+\sqrt{\frac{\omega^2}{4\eta_b^2}-1}\right), & \text{if } \omega < -2\eta_b\\
\frac{1}{\eta_b}\left(\frac{\omega}{2\eta_b}-i\sqrt{1-\frac{\omega^2}{4\eta_b^2}}\right), & \text{if } \abs{\omega} < 2\eta_b\label{Sigma(omega)}.
\end{cases}
\end{align}
Using these results,  the terms involved in the NEGF-expression for conductance become
\begin{align}
T_1(\omega)=4\eta_c^4g^2(\omega)\abs{[G_1^+(\omega)]_{1N}}^2, \label{st1}\\
T_2(\omega)=4\eta_c^4g^2(\omega)\abs{[G_2^+(\omega)]_{1N}}^2, \label{st2}\\
T_3(\omega)=4\eta_c^4g^2(\omega)\abs{[G_2^+(\omega)]_{11}}^2. \label{st3}
\end{align}
We  will use these expressions  for the analytical proof of the equivalence of the two methods in Sec.~\ref{sec_proof}.

 We now have the conductance at the left junction from the QLE-NEGF approach and we want to compare these results with the results from the scattering approach to the same problem. Here we present a qualitative discussion  of the scattering formalism and relegate the details of the calculation to Sec.~(\ref{scattering}). The first step in the scattering approach would be to identify the different scattering processes that could take place in the system. Let us consider a plane wave incident on the wire from the left reservoir and then, considering the wire as a scatterer, we note that there are  a total of four processes that can take place. Two of these processes are ---  (i) an electron from the left reservoir being reflected back into the left reservoir and (ii) an electron from the left reservoir being transmitted across the wire into the right reservoir. We will refer to these as normal reflection and normal transmission processes respectively. However, in a superconductor the electron and hole wavefunctions are intertwined  and therefore an electron may get scattered as a hole also. This results in the two additional scattering process in which  an electron from the left reservoir can (iii) get reflected back as a hole into the left reservoir  or (iv) get transmitted across the wire as a hole into the right reservoir. We refer to these as Andreev reflection and transmission respectively. During these two processes,  charge conservation is ensured by the formation of  a cooper pair  in the wire. 

Having identified the scattering processes, the next step would be to write down a stationary state wavefunction, at some energy $E$, in the three regions of the system (the wire, left bath, right bath) with appropriate scattering amplitudes and  wavefunctions so that all the scattering processes are captured. In the left reservoir, we thus have the incoming plane wave  and the outgoing plane waves for the reflected electron and hole corresponding to the normal and Andreev reflection respectively. The reflected electron and hole plane waves are multiplied by  some scattering amplitudes which we take to be $r_n$ and $r_a$ respectively. Similarly, in the right reservoir we will have the transmitted electron and hole plane waves from the normal transmission and the Andreev transmission respectively and we take the scattering amplitudes for these to be  $t_n$ and $t_a$ respectively. In the wire, the wavefunction will be a superposition of quasi particles of the wire at energy $E$ which are defined in terms of the  diagonalization of the bulk wire Hamiltonian. The normal and Andreev scattering amplitudes are  obtained by implementing the boundary conditions and then the conductance at the left junction, in units of $e^2/h$, is  given by the net probability of an electron to be transmitted across the left junction which is
\begin{equation}
G_L^S= \abs{t_n}^2+\abs{t_a}^2+2\abs{r_a}^2\label{GLscat1}=1-\abs{r_n}^2+\abs{r_a}^2.
\end{equation}
The last step follows from the probability conservation, $\abs{r_n}^2+\abs{r_a}^2+\abs{t_n}^2+\abs{t_a}^2=1$.  The factor $2$ in Eq.~(\ref{GLscat1}) with $|r_a|^2$ is due to the fact that in the Andreev reflection process, two electrons are transmitted across the junction as a single cooper pair. 

Now in order to compare these two independent approaches note that the NEGF expression for the current, Eq.~(\ref{currexp1}), has contribution from three terms.  On comparison of these three terms with the usual Landauer formulas for current one may expect the following:  the first term  has electrons  as incoming and  outgoing particles and therefore this must be the contribution of the electron from the left bath being scattered as an electron into the right bath (normal transmission), the second term having electrons and holes in the opposite baths as the incoming and outgoing particles respectively   should correspond to the process of an electron from the left bath being scattered as a hole into the  right bath (Andreev transmission). Finally, the third term which also has electrons and holes  as incoming and outgoing particles respectively but in the same bath would therefore correspond to the electron from the left bath scattered back as a hole into left bath again (Andreev reflection). The traces in the three terms should then be proportional to the probability of these three processes respectively. Therefore, the first two  terms of the conductance expression in Eq.~({\ref{GLnegf1}}) calculated at energy $E$, $T_1(E)$ and $T_2(E)$ should be equal to the probabilities from the scattering amplitudes $t_n$ and $t_a$ at the same energy respectively and the sum of the last two terms, $T_3(E)$ and $T_3(-E)$, both of which follow from the third term of the current expression in Eq.~(\ref{currexp1}) should then be equal to $2|r_a|^2$. This would make the two conductance expressions, in  Eq.~(\ref{GLnegf1}) and Eq.~(\ref{GLscat1}), from  the two approaches  exactly the same.  In Sec.~\ref{sec_proof} we present an exact proof of this result but for now we proceed to  Sec.~(\ref{scattering}) where we present the details of the calculations involved in  the scattering approach.

\section{Scattering approach}
\label{scattering}
  In this section, we first find out the stationary states of energy $E$ inside the left reservoir, the wire and the right reservoir. This would  enable us to write down the scattering wavefunction as discussed in Sec.~\ref{model} in the three regions and after implementing the boundary conditions, at the reservoir-wire junctions, we would obtain a set of linear equations for the normal and Andreev scattering amplitudes. The conductance could then be obtained via Eq.~(\ref{GLscat1}). Afterwards, we discuss the case of $E=0$ separately and find  the wavefunctions  and the parameter regime of existence of the  MBS.
  
  Consider for the moment the case where the wire has $N$ sites while the left and right reservoirs have $N_L$ and $N_R$ number of sites respectively so that the total number of sites is $N_S=N+N_L+N_R$. Let us define a column vector $\chi_p=\begin{pmatrix}  c_p \\ c_p^\dagger \end{pmatrix}$, where the index $p$ refers to any site on the entire system so that we can rewrite the Hamiltonian in  the form
  \begin{equation}
  H=\frac{1}{2}\sum_{p,q}\chi_p^\dagger\mathcal{A}_{pq}\chi_q\label{H2}
  \end{equation}
  where $\mathcal{A}_{pq}$ are $2\cross2$ block matrices which form the elements of the $2 N_S \times 2 N_S$  matrix $\mathcal{A}$ given by
 \begin{widetext}
\begin{equation} \label{Amatrix}
\mathcal{A}=\begin{pmatrix}
\mathbf{0}& A_L & & \\
A_L^T&\mathbf{0}& A_L & &\\
&& \ddots &&&&&& \\
&& A_L^T &\mathbf{0}&A_C&&&&\\
&&& A_C^T &A&A_W&&&&\\
&&&& A^T_W&A&A_W&&\\
&&&&&& \ddots &&&&& \\
&&&&&& A^T_W &A&A_C&&&\\
&&&&&&& A_C^T &\mathbf{0}&A_R\\
&&&&&&&& A_R^T&\mathbf{0}&A_R\\
&&&&&&&&&& \ddots & \\
&&&&&&&&&& A_R^T &\mathbf{0}
\end{pmatrix},
\end{equation}
\end{widetext}
with
\begin{align}
 A_{R}=&A_{L}= \begin{pmatrix}
 -\eta_b && 0\\
 0 &&\eta_b
\end{pmatrix}~~\text{,}~~&A_C= \begin{pmatrix}
-\eta_c && 0\\ 0 &&\eta_c
  \end{pmatrix}\text{,}\\ \label{Adef} A_W&= \begin{pmatrix}
  -\eta_w && -\Delta\\
  \Delta &&\eta_w
  \end{pmatrix}~~\text{,}~~&A= \begin{pmatrix}
  -\mu_w && 0\\
  0 &&\mu_w
  \end{pmatrix}. 
\end{align}
Now  considering first the wire region, let $\Psi_W(j)$ be the components of the stationary state of energy $E$ of the wire in this basis. Then, in the bulk of the wire we have:
\begin{equation}
A_W^T \Psi_W(j-1)+A\Psi_W(j)+A_W\Psi_W(j+1)=E\Psi_W(j). \label{psiw}
\end{equation}
We choose  $\Psi_W(j)=\begin{pmatrix}U\\V\end{pmatrix}z^{j}$ and fix $z$ such that the Eq.~(\ref{psiw}) is satisfied. On substitution of  $\Psi_W(j)$ in Eq.~(\ref{psiw}), we arrive at the following equation,

\begin{equation}
\begin{pmatrix}
\eta_w(z+\frac{1}{z})+\mu_w+E && \Delta(z-\frac{1}{z})\\
\Delta(z-\frac{1}{z}) && \eta_w(z+\frac{1}{z})+\mu_w-E
\end{pmatrix}\begin{pmatrix}
U\\V
\end{pmatrix}=0, \label{UVeq}
\end{equation}
which means that $z$ must be fixed such that
\begin{equation}
\begin{vmatrix}
\eta_w(z+\frac{1}{z})+\mu_w+E && \Delta(z-\frac{1}{z}) \\
\Delta(z-\frac{1}{z}) && \eta_w(z+\frac{1}{z})+\mu_w-E
\end{vmatrix}=0.
\end{equation}
Clearly, there are four possible solutions for  $z$ as this determinant on expansion will give a fourth order equation in $z$.  However, we can make things a bit simpler by choosing $z=e^{x}$ so that the above determinant on expansion gives a quadratic equation in $\cosh x$ rather than a fourth order equation in $z$. The quadratic equation thus obtained is the following:
\begin{equation}
(\cosh x)^2-\frac{\mu\eta_w}{\Delta^2-\eta_w^2}\cosh x+\frac{E^2-\mu^2-4\Delta^2}{4(\Delta^2-\eta_w^2)}=0
\end{equation}   
with its  two solutions  given by
\begin{equation}
\cosh x_\pm = \frac{\mu_w\eta_w \pm\sqrt{(\eta_w^2-\Delta^2)(E^2-4\Delta^2)+\Delta^2\mu_w^2}}{2(\Delta^2-\eta_w^2)}. \label{coshxpm}
\end{equation}
Therefore, the four possible solutions to $z$, which are obtained from the two quadratic equations $z^2-2\cosh x_\pm z +1=0$, are given by
\begin{align}
	z_1=e^{-x_+},~ z_2=e^{-x_-},~z_3=e^{x_+},~z_4=e^{x_-}.
\end{align}
From Eq.~(\ref{UVeq}), we see that $U$ and $V$ for the corresponding solutions for $z$ could be chosen in the following form:
\begin{align}
U_s&=-\Delta(z^2_s-1)\\
V_s&=\eta_w(z_s^2+1)+z_s(\mu_w+E)
\end{align}
where $s=1,2,3,4$ for the four solutions of $z$. Therefore, we have the required stationary states inside the wire. Now, the stationary states of energy $E$ inside the baths can be obtained from the wire solution via the transformation,  $\mu_w\rightarrow0$, $\Delta\rightarrow0$ and $\eta_w\rightarrow\eta_b$. We get the solutions to be two left travelling plane waves and two right travelling plane waves of the following forms:
\begin{align}
	\begin{pmatrix}1\\0\end{pmatrix}e^{iqx}, ~~~ \begin{pmatrix}1\\0\end{pmatrix}e^{-iqx},\label{pwelec}\\
	\begin{pmatrix}0\\1\end{pmatrix}e^{iq'x}, ~~~ \begin{pmatrix}0\\1\end{pmatrix}e^{-iq'x}, \label{pwhole}
\end{align}
where $q$ and $q'$ are given by $\cos^{-1}\left(-\frac{E}{2\eta_b}\right)$ and $\cos^{-1}\left(-\frac{E}{2\eta_b}\right)-\pi$ respectively. Physically, the first two solutions, Eq.~(\ref{pwelec}), correspond to an electron travelling right and left respectively while the last two solutions, Eq.~(\ref{pwhole}), correspond to a hole travelling to the right and left respectively. %These two solutions are obtained by defining $U$ and $V$ from the second component equation of Eq.~(\ref{UVeq}). 

We are now in a position to write the explicit form of the scattering wavefunction  in the three regions for a plane wave of energy $E$ incident from the left reservoir. This will be of the form:
\begin{align}
&\Psi_L(\alpha)=\begin{pmatrix}1 \\ 0\end{pmatrix}e^{iq\alpha}+r_n\begin{pmatrix}1 \\ 0\end{pmatrix}e^{-iq\alpha}+r_a\begin{pmatrix}0 \\ 1	\end{pmatrix}e^{-iq^\prime\alpha}\label{psil}\\
&\Psi_W(j)=\sum_{s=1}^4 a_s\begin{pmatrix}U_s \\ V_s\end{pmatrix}z_s^{j-1}\label{pssiw}\\%+a_2\begin{pmatrix}U_2 \\ V_2\end{pmatrix}z_2^{j-1}+a_3\begin{pmatrix}U_3 \\ V_3\end{pmatrix}z_3^{j-1}+a_4\begin{pmatrix}U_4 \\ V_4\end{pmatrix}z_4^{j-1}\\
&\Psi_R(\alpha^\prime)=t_n\begin{pmatrix}1 \\ 0\end{pmatrix}e^{iq(\alpha^\prime-N-1)}+t_a\begin{pmatrix}0 \\ 1	\end{pmatrix}e^{iq^\prime(\alpha^\prime-N-1)}\label{psir}
\end{align}
with $\alpha=-\infty,...,-1,0 ~~,~~j=1,2,...N ~\text{and}~\alpha^\prime=N+1,N+2,...\infty$. As already discussed in  Sec.~(\ref{model}), $r_n$ is the probability amplitude for the electron to be reflected back at the left junction as an electron. Therefore, this corresponds to the normal reflection. $r_a$ is the probability amplitude for the  Andreev reflection. Similarly, $t_n$ and $t_a$ are the normal and the Andreev transmission amplitudes respectively. The solution inside the wire represents a superposition, with amplitudes $a_1,a_2,a_3$ and $a_4$, of the quasi-particles with energy $E$ in the wire travelling to the left and right respectively.  These scattering amplitudes are obtained by implementing the boundary conditions. We note that we have eight scattering amplitudes and two boundaries, one at the left end and the other at the right end of the wire. Each  
 site on either side of each boundary gives two equations. Therefore, a single boundary gives four equations in total and we have exactly eight equations from the two boundaries, sufficient to determine the eight scattering  amplitudes. These eight boundary equations are given by
 \begin{align}
 	&A_L\Psi_L(-1)+A_C\Psi_W(1)=E\Psi_L(0),\label{bcl1}\\
 	&A_C\Psi_L(0)+A\Psi_W(1)+A_W\Psi_W(2)=E\Psi_W(1),\label{bcl2}\\
 	&A_W^T\Psi_W(N-1)+A\Psi_W(N)+A_C\Psi_R(N+1)=E\Psi_W(N),\label{bcr2}\\
 	&A_C^T\Psi_W(N)+A_R\Psi_R(N+2)=E\Psi_R(N+1).\label{bcr1}
 \end{align}     
 After substituting the solution from Eqs.~(\ref{psil}, \ref{pssiw}, \ref{psir}), the eight linear equations for the scattering amplitudes can be expressed in matrix form as
 \begin{widetext}

 	\begin{equation}\label{ampeqns}\begingroup % keep the change local
 	\setlength\arraycolsep{1pt}
 	\begin{pmatrix}
 	\eta_b e^{iq}+E && 0 &&\eta_c U_{1}&&\eta_c U_2&&\eta_c U_{3}&&\eta_c U_{4}&&0&&0\\
 	0&& \eta_b e^{iq^\prime}-E &&\eta_c V_{1}&&\eta_c V_{2}&&\eta_c V_{3}&&\eta_c V_{4}&&0&&0\\
 	\eta_c &&0 && f_{1} &&f_{2}&&f_{3}&&f_{4}&&0&&0\\
 	0   &&\eta_c && f^\prime_{1} &&f^\prime_{2}&&f^\prime_{3}&&f^\prime_{4}&&0&&0\\
 	0 &&0 && g_{1} &&g_{2}&&g_{3}&&g_{4}&&\eta_c&&0\\
 	0   &&0 && g_{1}^\prime &&g_{2}^\prime&&g_{3}^\prime&&g_{4}^\prime&&0&&\eta_c\\
 	0 && 0 &&\eta_c z_{1}^{N-1}U_{1}&&\eta_cz_{2}^{N-1} U_{2}&&\eta_cz_{3}^{N-1} U_{3}&&\eta_cz_{4}^{N-1} U_{4}&&\eta_be^{iq}+E&&0\\
 	0 && 0 &&\eta_c z_{1}^{N-1}V_{1}&&\eta_cz_{2}^{N-1} V_{2}&&\eta_cz_{3}^{N-1} V_{3}&&\eta_cz_{4}^{N-1} V_{4}&&0&&\eta_be^{iq^\prime}-E	
 	\end{pmatrix}\begin{pmatrix}
 	r_n\\r_a\\a_{1}\\a_{2}\\a_{3}\\a_{4}\\t_n\\t_a
 	\end{pmatrix}= \begin{pmatrix}
 	-\eta_b e^{-iq}-E\\0\\-\eta_c\\0\\0\\0\\0\\0
 	\end{pmatrix},
 	\endgroup
 	\end{equation}
\end{widetext}
 where for $s=1,2,3,4$ the $f_s$, $f_s'$, $g_s$ and $g_s'$ are given by 
\begin{align}
f_s&=(\mu_w+E)U_s+z_s(\eta_w U_s +\Delta V_s),\\
f^\prime_s&=(\mu_w-E)V_s+z_s(\eta_w V_s +\Delta U_s),
\end{align}
\begin{align}
g_s&=(\mu_w+E)U_sz_s^{N-1}+(\eta_wU_s-\Delta V_s) z_s^{N-2},\\
g_s^\prime&=(\mu_w-E)V_sz_s^{N-1}+(\eta_wV_s-\Delta U_s) z_s^{N-2}.
\end{align}
Solving Eqs.~(\ref{ampeqns}) gives us the required expressions for  $r_n$, $r_a$, $t_n$ and $t_a$, and from these we can obtain the conductance using the scattering approach. 

 We now  look for the special solution corresponding to the zero energy MBS in this open wire system. We expect that for long enough wires there are two MBS each localized at edges of the wire. Let us consider a zero energy eigenstate localized at the left end $(j=1)$, therefore for this we must have $t_n=0$, $t_a=0$ and $a_s=0$ if $|z_s|\geq 1$ in the wavefunction given by Eq.~(\ref{psil}, \ref{pssiw}, \ref{psir}) with $E=0$. Out of  the four roots, $z_1$, $z_2$, $z_3$ and $z_4$, it is clear that two of them  always have absolute values greater than $1$ while the other two are always less that $1$. Let us choose, by relabelling, $z_1$ and $z_2$ to be the ones with absolute values  less  than $1$, therefore we set $a_3$ and $a_4$ to be zero. We also note from Eq.~(\ref{UVeq}) that for $E=0$, $U=\pm V$.  Therefore, depending on whether $z_1(z_2)$ satisfies $U_1=V_1(U_2=V_2)$ or $U_1=-V_1(U_2=-V_2)$ we choose $U_1(U_2)$ and $V_1(V_2)$ accordingly. This choice could be made by noting that $U=V$ is satisfied by
  \begin{align}
  	z_\pm=\frac{-\mu_w\pm\sqrt{\mu_w^2-4\eta_w^2+4\Delta^2}}{2(\eta_w+\Delta)},
  \end{align}
  while $U=-V$ is satisfied by
  \begin{align}
  z_\pm'=\frac{-\mu_w\pm\sqrt{\mu_w^2-4\eta_w^2+4\Delta^2}}{2(\eta_w-\Delta)}.
  \end{align}
  Thus $z_1$ and $z_2$ have to be equal to two of these four roots which have absolute values less than $1$.   Fixing  $\Delta>0$, we find that for $|\mu_w|<2|\eta_w|$, $|z_\pm|<1$  and $|z_{\pm}'|>1$ while for $|\mu_w| > 2|\eta_w|$, the absolute value  one of the roots among $z_\pm$ and $z_\pm'$ is greater than 1 while the other is less than 1. This implies that for $\Delta>0$ and $|\mu_w|<2|\eta_w|$, we need to set  $U_1=V_1$ and $U_2=V_2$,  while for $\Delta>0$ and $|\mu_w| > 2|\eta_w|$, we have $U_1=V_1$ and $U_2=-V_2$.
 
 We first take the case with $\Delta>0$ and $|\mu_w|<2|\eta_w|$. For this we have   $t_n=t_a=a_3=a_4=0$, $E=0$, $U_2=V_1$ and $U_2=V_2$, which simplify Eq.~(\ref{ampeqns})  to the set of four equations:  
 \begin{align}
 	&i\eta_br_n+\eta_cU_1 a_1 +\eta_c U_2 a_2=i\eta_b,\\
	-&i\eta_br_a+\eta_cU_1 a_1 +\eta_c U_2 a_2=0,\\
 	&\eta_cr_n+\kappa_1U_1 a_1 +\kappa_2 U_2 a_2=-\eta_c,\\
 	&\eta_cr_a+\kappa_1U_1 a_1 +\kappa_2 U_2 a_2=0,
 \end{align}
 where $\kappa_s=\mu_w+z_s(\eta_w+\Delta)$. These equations  can  be solved to give $r_n=0$, $r_a=1$,
 \begin{equation}
 	U_1a_1=\frac{(i\eta_b\kappa_2+\eta_c^2)}{\eta_c(\kappa_2-\kappa_1)}~\text{and}~ U_2a_2=-\frac{(i\eta_b\kappa_1+\eta_c^2)}{\eta_c(\kappa_2-\kappa_1)}.
 \end{equation}
These equations then give the wavefunction of the zero mode  that is localized at the left end  and can be written as
 \begin{align}
 &\Psi_L^{MBS}(\alpha)=\begin{pmatrix} 1\\1 \end{pmatrix}\sin \frac{\pi \alpha}{2}~\text{and}~\Psi_R^{MBS}(\alpha^\prime)=0\label{mbspsir},\\
 &\Psi_W^{MBS}(j)=\begin{pmatrix}1 \\ 1\end{pmatrix}\Im\left[\frac{(i\eta_b\kappa_2+\eta_c^2)z_1^{j-1}-(i\eta_b\kappa_1+\eta_c^2)z_2^{j-1}}{\eta_c(\kappa_2-\kappa_1)}\right].\label{mbspssiw}
 \end{align}
Also, due to the perfect Andreev reflection($r_a=1$), we get $G_L(E=0)=2$ which marks the zero bias peak found in systems which host MBS\cite{roy2012,mourik2012,das2012zero}. Thus the zero mode found is the wavefunction of the zero energy MBS found to be present in the parameter regime $|\mu_w|<2|\eta_w|$. Due to the left-right symmetry of the Hamiltonian, the wavefunction of the MBS localized at the other end of the wire can directly be written as:
\begin{align}
&\Phi_R^{MBS}(\alpha')=\begin{pmatrix} 1\\1 \end{pmatrix}\sin \frac{\pi (N-\alpha')}{2}~\text{and}~\Phi_L^{MBS}(\alpha)=0, \label{mbs1psir}\\
&\Phi_W^{MBS}(j)=\begin{pmatrix}1 \\ 1\end{pmatrix}\Im\left[\frac{(i\eta_b\kappa_2+\eta_c^2)z_1^{N-j}-(i\eta_b\kappa_1+\eta_c^2)z_2^{N-j}}{\eta_c(\kappa_2-\kappa_1)}\right].\label{mbs1pssiw}
\end{align}
 \begin{figure}[htb!]
	\centering
	\subfigure[]{
		\includegraphics[width=8.cm,height=60mm]{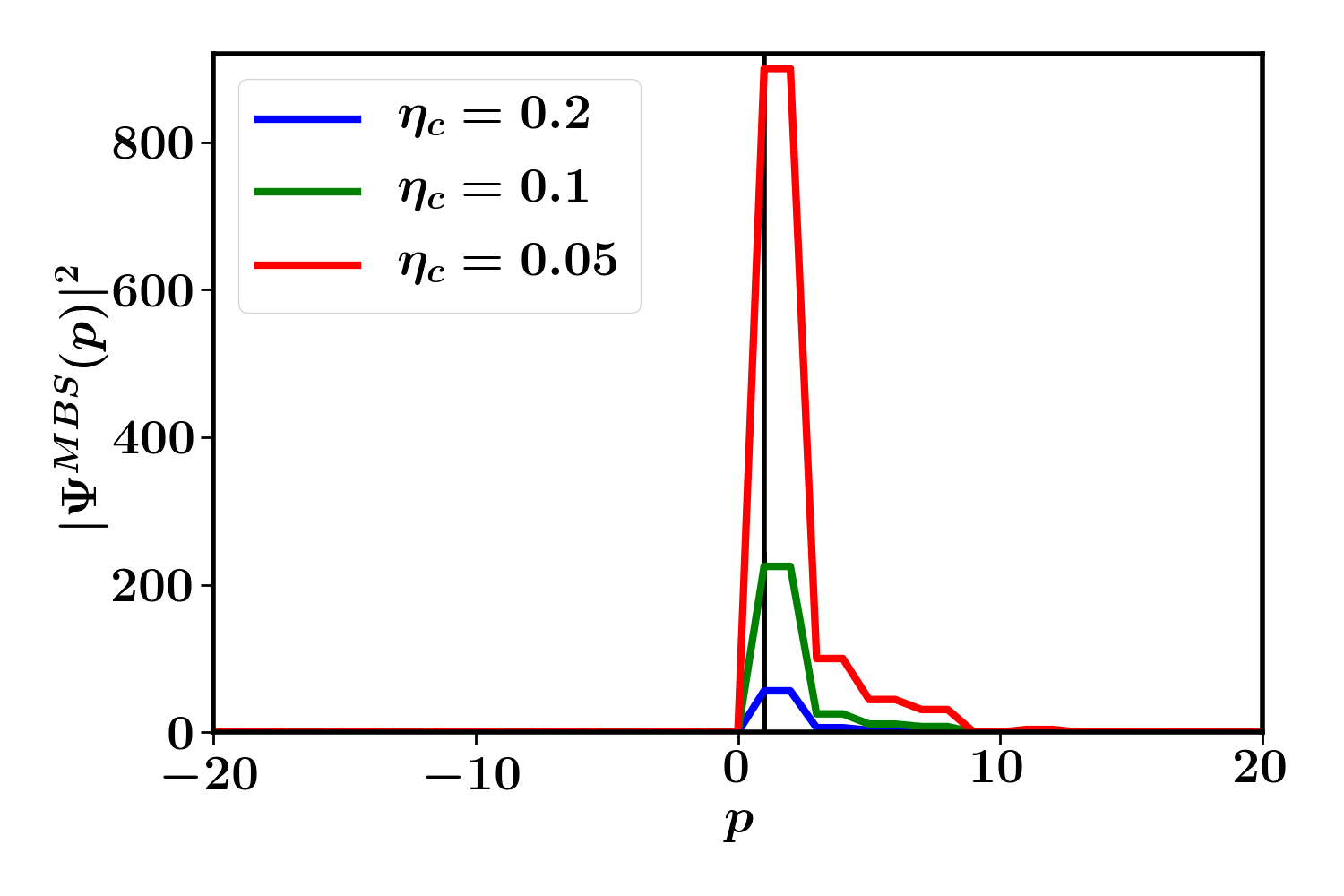}
	}
\caption{ Plot of the MBS wavefunction for different couplings with the reservoir at parameter values-- $\eta_b=1.5$, $\mu_w=0.5$, $\Delta=0.5$, $\eta_w=1$.  The normalization of these wavefunctions is the same as in Eq.~(\ref{mbspsir}-\ref{mbspssiw}) and the vertical black line marks the left end of the wire.  Note that the lead wavefunctions are not visible on this scale.}
\label{etacmbs}
\end{figure}
  The absolute value of  the height of the peak in the MBS wavefunction is given by $|\frac{\eta_b}{\eta_c}|$. Therefore, the height of the peak decreases as coupling with the bath increases which makes sense since one expects the wavefunction to leak into the reservoir more as the coupling with reservoirs increases. This can be seen   in Fig.~\ref{etacmbs} where we plot the MBS wavefunction for a few different couplings with the reservoirs. Also, increasing $\eta_b$ increases the band width of the system which decreases the density of the states around $E=0$ and therefore the MBS of the isolated wire hybridizes less with the reservoir wavefunctions as the energy difference between them increases. Note that if the height of the peak in the MBS goes down, the weight of the MBS in the reservoirs increases and vice-versa. We will  see later that this wavefunction helps in explaining the behaviour of the zero bias peak with different parameters of the Hamiltonian. 
  
  Let us consider the case with $\Delta>0$ and $|\mu_w|>2|\eta_w|$. For this we have  
   \begin{align}
  &i\eta_br_n+\eta_cU_1 a_1 +\eta_c U_2 a_2=i\eta_b,\\
  -&i\eta_br_a+\eta_cU_1 a_1 -\eta_c U_2 a_2=0,\\
  &\eta_cr_n+\kappa_1U_1 a_1 +\kappa_1 U_2 a_2=-\eta_c,\\
  &\eta_cr_a+\kappa_1U_1 a_1 -\kappa_1 U_2 a_2=0.
  \end{align}
   These equations can be solved for $r_n$, $r_a$, $U_1$ and $U_2$ with which we can then construct the zero mode present in this parameter regime. However, these equations give $r_a=0$ and therefore there is no perfect Andreev reflection ($r_a=1$). Thus  the zero mode constructed out of them would not be the MBS. They would merely be the zero energy states of the left reservoir leaking into the wire.  We therefore conclude that only the zero energy states present in the parameter regime $|\mu_w|<2|\eta_w|$ give rise to the perfect Andreev reflection and are the states representing the MBS of this system. Similar arguments  can be repeated for the case $\Delta<0$.

\section{Analytical proof of the equivalence of QLE-NEGF and Scattering approaches}
\label{sec_proof}
In this section we will show analytically the equivalence between the two approaches by deriving the following equalities,   
\begin{align}
	T_1(E)=|t_n|^2,&~T_2(E)=|t_a|^2\\
	\text{and}~T_3(E)=&T_3(-E)=|r_a|^2
\end{align}
where, $T_1(E)$, $T_2(E)$ and $T_3(E)$  are given Eq.~(\ref{st1}, Eq.~(\ref{st2}) and Eq.~(\ref{st3}) respectively with  $\mu_L$ replaced by $E$. This would then straight forwardly imply the equivalence of the two conductance expression. To proceed, we first need to find a set of equations relating the transmission amplitudes, $t_n$ and $t_a$, to the reflection amplitudes, $r_n$ and $r_a$ directly, which is possible by relating $\begin{pmatrix}
\Psi_L(-1)\\\Psi_L(0)
\end{pmatrix}$ directly to $ \begin{pmatrix}
\Psi_R(N+1)\\\Psi_R(N+2)
\end{pmatrix}$ via  transfer matrices. We start by considering the equation for the stationary state of energy $E$ inside the wire
\begin{equation}
	A_W^T \Psi_W(j-1)+A\Psi_W(j)+A_W\Psi_W(j+1)=E\Psi_W(j)
\end{equation}
which we re-write  in the following recursive form:
\begin{eqnarray}
	\begin{pmatrix}
		A_W^T\Psi_W(j-1)\\\Psi_W(j)
	\end{pmatrix}=\Omega_W\begin{pmatrix}
		A_W^T\Psi_W(j)\\\Psi_W(j+1)
	\end{pmatrix}, \label{tmeq}
\end{eqnarray}
where
\begin{align}
	\Omega_W=\begin{pmatrix}
		(E-A) A_W^{-T} && -A_W\\
		A_W^{-T} && 0
	\end{pmatrix}. \label{OmegaW}
\end{align}
Using the   boundary  conditions at the left junction, Eq.~(\ref{bcl1}) and Eq.~(\ref{bcl2}),  we can write
\begin{equation}
	\begin{pmatrix}
		\Psi_L(-1)\\\Psi_L(0)
	\end{pmatrix}=\Omega_{L1}\Omega_{L2}\begin{pmatrix}
		A_W^T\Psi_W(1)\\\Psi_W(2)
	\end{pmatrix},\label{rec1}
\end{equation}
where
\begin{align}
	\Omega_{L1}=\begin{pmatrix}
		A_L^{-1}EA_C^{-1} && -A_L^{-1}A_C\\ A_C^{-1} && 0
	\end{pmatrix},\\\label{OmegaL2}\Omega_{L2}=\Omega_W=\begin{pmatrix}
		(E-A)A_W^{-T} && -A_W\\ A_W^{-T} &&0
	\end{pmatrix}.%=\begin{pmatrix}EA_c^{-1}(E-A)A_w^{-T}-A_cA_w^{-T} && -EA_c^{-1}A_w\\ A_c^{-1}(E-A)A_w^{-T} && -A_c^{-1} A_w	\end{pmatrix}
\end{align}
Using Eq.~(\ref{tmeq})  repeatedly in Eq.~(\ref{rec1}) we have the following equation:
\begin{equation}
	\begin{pmatrix}
		\Psi_L(-1)\\\Psi_L(0)
	\end{pmatrix}=\Omega_{L1}\Omega_{L2}\Omega^{N-2}_W\begin{pmatrix}
		A_W^T\Psi_W(N-1)\\\Psi_W(N)
	\end{pmatrix}.
\end{equation}
Finally, we use the  boundary conditions at the right junction, Eq.~(\ref{bcr2}) and Eq.~(\ref{bcr1}), to obtain the desired equation
\begin{align}
	\begin{pmatrix}
		\Psi_L(-1)\\\Psi_L(0)
	\end{pmatrix}&=\Omega_{L1}\Omega_{L2}\Omega^{N-2}_W\Omega_{R2}\Omega_{R1}\begin{pmatrix}
		\Psi_R(N+1)\\\Psi_R(N+2)
	\end{pmatrix}\\	
	&=\Omega_{L1}\Omega\Omega_{R1}\begin{pmatrix}
		\Psi_R(N+1)\\\Psi_R(N+2)
	\end{pmatrix},\label{tmeqt}
\end{align}
where
\begin{align}
	\Omega_{R2}&=\begin{pmatrix}
		E-A && -I \\ I&&0
	\end{pmatrix},\label{OmegaR2}\\
	\Omega_{R1}&=\begin{pmatrix}
		A_C^{-T}E && -A_C^{-T}A_R\\
		A_C &&0
	\end{pmatrix},\\
\Omega&=\Omega_{L2}\Omega^{N-2}_W\Omega_{R2},
\end{align}
 and $I$ denotes a $2 \times 2$ unit matrix. 
 We now have Eq.~(\ref{tmeqt}) which relates $\begin{pmatrix}
\Psi_L(-1)\\\Psi_L(0)
\end{pmatrix}$ directly to $ \begin{pmatrix}
\Psi_R(N+1)\\\Psi_R(N+2)
\end{pmatrix}$  via the transfer matrix, $\Omega_{L1}\Omega\Omega_{R1}$. This equation will furnish a set of four equations for $r_n$, $r_a$, $t_n$ and $t_a$ after using the forms of $\Psi_L(\alpha)$ and $\Psi_R(\alpha')$ from Eq.~(\ref{psil}) and Eq.~(\ref{psir}) respectively. However, we could make things much more simpler by using the forms of the matrices $A_C$, $A_L$ and $A_R$ to write 

\begin{align}
	\Omega_{L1}&=\frac{1}{\eta_b\eta_c}\begin{pmatrix}E && -\eta_c^2 \\ -\eta_b\sigma^z &&0 \end{pmatrix},\\
	\Omega_{L1}^{-1}&=\frac{1}{\eta_c}\begin{pmatrix}
		0 && -\eta_c^2\sigma^z\\ -\eta_b && -E\sigma^z
	\end{pmatrix},\\
	\text{and}\hspace{0.5cm}\Omega_{R1}&=\frac{1}{\eta_c}\begin{pmatrix}
		-E\sigma^z && -\eta_b\\
		-\eta_c^2\sigma^z && 0
	\end{pmatrix},
\end{align}
where $\sigma^z=\begin{pmatrix}
1 &&0\\
0&&-1
\end{pmatrix}$. Now, from Eq.~(\ref{tmeqt}) we have
\begin{equation}
	\Omega_{L1}^{-1}\begin{pmatrix}\Psi_L(-1)\\\Psi_L(0)
	\end{pmatrix} =\Omega\Omega_{R1}\begin{pmatrix}
		\Psi_R(N+1)\\\Psi_R(N+2)
	\end{pmatrix},
\end{equation}
which then gives the following two matrix equations:
\begin{align}
	&\eta_c^2\sigma^z\Psi_L(0)=\bar\Omega_{11}[E\sigma^z\Psi_R(N+1)+\eta_b\Psi_R(N+2)]\notag\\&\hspace{3cm}+\bar\Omega_{12}\eta_c^2\sigma^z\Psi_R(N+1),\label{PsiL(0)}\\
	&\eta_b\Psi_L(-1)+E\sigma^z\Psi_L(0)=\bar\Omega_{21}E\sigma^z\Psi_R(N+1)\notag\\&\hspace{2cm}+\bar\Omega_{21}\eta_b\Psi_R(N+2)+\bar\Omega_{22}\eta_c^2\sigma^z\Psi_R(N+1), \label{PsiL(-1)}
\end{align}
where $\bar{\Omega}_{ij}$ are $2\cross2$  matrices that form blocks of the matrix $\Omega$, \emph{i.e}
\begin{equation}
\Omega=	\begin{pmatrix}\bar{\Omega}_{11}&& \bar{\Omega}_{12}\\\bar{\Omega}_{21}&&\bar{\Omega}_{22}\end{pmatrix}.
\end{equation}
Using the forms of $\Psi_L(\alpha)$ and $\Psi_R(\alpha')$ from Eq.~(\ref{psil}) and Eq.~(\ref{psir}) respectively, Eq.~(\ref{PsiL(0)}) and Eq.~(\ref{PsiL(-1)}) can be written  as
\begin{align}
	&\eta_c^2(\ket{+}+r_n\ket{+}-r_a\ket{-})=\notag\\&\hspace{1cm}\left[-\eta_be^{-iq}\bar{\Omega}_{11}+\eta_c^2\bar{\Omega}_{12}\right][t_n\ket{+}-t_a\ket{-}],\label{4set1}\\	
	-&\eta_b(e^{iq}\ket{+}+e^{-iq}r_n\ket{+}-e^{-iq}r_a\ket{-})=\notag\\&\hspace{1cm}\left[-\eta_be^{-iq}\bar{\Omega}_{21}+\eta_c^2\bar{\Omega}_{22}\right][t_n\ket{+}-t_a\ket{-}],\label{4set2}
\end{align}
where we substituted $q-\pi$ for $q'$, $\ket{\pm}$  is the eigenvector of $\sigma^z$ with eigenvalue $\pm1$. We can simultaneously get rid of  $r_n$ and $r_a$ by subtracting  Eq.~(\ref{4set1}) and Eq.~(\ref{4set2}) after multiplication with appropriate factors. Thus, one finds:
\begin{equation}
	-2ie^{iq}\sin q \ket{+}=\frac{\eta_b}{\eta_c^2}\mathcal{O}[t_n\ket{+}-t_a\ket{-}],\label{stntaeq}
\end{equation}
where
\begin{equation}
	\mathcal{O}=\left[-e^{-2iq}\bar{\Omega}_{11}+\frac{\eta_c^2}{\eta_b}e^{-iq}\bar{\Omega}_{12}-\frac{\eta_c^2}{\eta_b}e^{-iq}\bar{\Omega}_{21}+\frac{\eta_c^4}{\eta_b^2}\bar{\Omega}_{22}\right]. \label{McalO}
\end{equation}
From Eq.~(\ref{stntaeq})  we can  write down the  two equations for $t_n$ and $t_a$:
\vspace{0.1cm}
\begin{align}
	1 =-\frac{t_n}{2i\frac{\eta_c^2}{\eta_b}\sin q}\bra{+}\mathcal{O}\ket{+}+\frac{t_a}{2i\frac{\eta_c^2}{\eta_b}\sin q}\bra{+}\mathcal{O}\ket{-}, \label{tntascateq1}\\	
	0=-\frac{t_n}{2i\frac{\eta_c^2}{\eta_b}\sin q}\bra{-}\mathcal{O}\ket{+}+\frac{t_a}{2i\frac{\eta_c^2}{\eta_b}\sin q}\bra{-}\mathcal{O}\ket{-}. \label{tntascateq2}
\end{align}
Also, from  Eq.~(\ref{4set2}) we directly get an expression of $r_a$ in terms of $t_n$ and $t_a$:
\begin{align}
	r_a&=\left[-\bra{-}\bar{\Omega}_{21}\ket{+}+e^{iq}\frac{\eta_c^2}{\eta_b}\bra{-}\bar{\Omega}_{22}\ket{+}\right]t_n\notag\\&~~-\left[-\bra{-}\bar{\Omega}_{21}\ket{-}+e^{iq}\frac{\eta_c^2}{\eta_b}\bra{-}\bar{\Omega}_{22}\ket{-}\right]t_a. \label{raexp}
\end{align}
For the moment we leave this here and turn our attention to the terms in the NEGF-expression for conductance. From Eq.~(\ref{st1}-\ref{st3}) we see that $T_1(E)$, $T_2(E)$ and $T_3(E)$ are  essentially given by the elements $[G_1^+(E)]_{1N}$, $[G_2^+(E)]_{1N}$ and $[G_2^+(E)]_{11}$ of the two Green's functions. We note that given the forms of the Green's functions $G_1^+(\omega)$ and $G_2^+(\omega)$, it is not easy to obtain these elements. Therefore, we have to re-write these Green's functions in some other form so that these elements could be obtained analytically. For that, we consider the Fourier transformed Langevin equations of motion for the wire, Eq.~(\ref{fts}), and  write its solution in a slightly different form involving a single $2N \times 2N$ Greens function. We start with the equations 
\begin{align}
	&[\Pi(\omega)]_{lm}\tilde{c}_m(\omega)-K_{lm}\tilde{c}_m^\dagger(-\omega)=\tilde{\eta}_l^L(\omega)+\tilde{\eta}_l^R(\omega), \label{fts1}\\
	&[\Pi(-\omega)]^*_{lm}\tilde{c}_m^\dagger(-\omega)-K_{lm}^*\tilde{c}_m(\omega)=\tilde{\eta}_l^{L\dagger}(-\omega)+\tilde{\eta}_l^{R\dagger}(-\omega). \label{fts2}
\end{align}
 Defining the two component vectors 
\begin{align}
	\notag C_i(\omega)=\begin{pmatrix}
		\tilde{c}_i(\omega)\\\tilde{c}^\dagger_i(-\omega)
	\end{pmatrix}~\text{and}~ \xi_i(\omega)=\begin{pmatrix}-\tilde\eta^L_i(\omega)-\tilde\eta^R_i(\omega)\\\tilde\eta_i^{L\dagger}(-\omega)+\tilde\eta_i^{R\dagger}(-\omega)\end{pmatrix},
\end{align}
we write  Eq.~(\ref{fts1}) and Eq.~(\ref{fts2}) together as $[\mathcal{G}^{-1}(\omega)]_{lm}C_m(\omega)=\xi_l(\omega)$ which has the solution:
\begin{align}
	C_l(\omega)=[\mathcal{G}(\omega)]_{lm}\xi_m(\omega)\label{sgfeq},
\end{align}	
with $\mathcal{G}^{-1}(\omega)$ being a $2N\cross2N$ matrix whose $lm$-th $2\cross2$ matrix block element is given by
\begin{equation}
	[\mathcal{G}^{-1}(\omega)]_{lm}=\begin{pmatrix}-[\Pi(\omega)]_{lm}&& K_{lm}\\-K_{lm}^*&&[\Pi(-\omega)]_{lm}^*\end{pmatrix}\label{mcG}
\end{equation}	
Comparing Eq.~(\ref{sgfeq}) with Eq.~(\ref{ngef}) for the $\tilde c_m(\omega)$ we see that
\begin{equation}
	[\mathcal{G}(\omega)]_{lm}=\begin{pmatrix} -[G_1^+(\omega)]_{lm} && [G_2^+(\omega)]_{lm}\\ -[G_2^+(-\omega)]^*_{lm}&& [G_1^+(-\omega)]_{lm}^*\end{pmatrix}.
\end{equation}

%This can also be shown explicitly by using the fact that $\mathcal{G}^{-1}\mathcal{G}=I$ and then using the forms of $\mathcal{G}$ and $\mathcal{G}^{-1}$ given by Eq.~(\ref{mcG}) and Eq.~(\ref{mcGinv}). 
Now, from Eq.~(\ref{Piij}), Eq.~(\ref{Klm}) and Eq.~(\ref{mcG}) we find that the matrix  $\mathcal{G}(E)$ has the following structure:

\begin{widetext}
	\begin{equation}
		\mathcal{G}(E)=\begin{pmatrix}
			-E+A-A_\Sigma  & A_W&0  &0& \dots\dots& 0&0 \\
			A_W^T & -E+A & A_W&0 & \dots\dots&0&0 \\
			0 & A_W^T & -E+A & A_W &\dots \dots &0&0\\
			\vdots&\vdots &\vdots & \ddots&&\vdots&\vdots \\
			\vdots&\vdots&\vdots& & \ddots &\vdots&\vdots \\
			\vdots&\vdots&\vdots&\dots& A_W^T&-E+A& A_W\\
			0&0 & 0&\dots&\dots&A_W^T &-E+A-A_\Sigma 
		\end{pmatrix}^{-1}\label{GME}
	\end{equation}
\end{widetext}

where $A_\Sigma=\begin{pmatrix}
-\eta_c^2 \Sigma(E)&&0\\
0&&\eta_c^2\Sigma^*(-E)
\end{pmatrix}$,   $\Sigma(E)$  being given by Eq.~(\ref{Sigma(omega)}) with $\mu_L$ replaced by $E$,  and the matrices $A,~A_W$  defined as in Eq.~\eqref{Adef}.  We note that for $|E|<2\eta_b$, $-\eta_b\Sigma(E)=\eta_b\Sigma^*(-E)=e^{iq}$. This then simplifies $A_\Sigma$ to be $\frac{\eta_c^2}{\eta_b}e^{iq}I_2$ with $I_2$ being a $2\cross2$ identity matrix. We work in the regime of $|E|<2\eta_b$ as outside of it the conductance is zero. Note that $\mathcal{G}(E)=\mathcal{G}^T(E)$ and therefore, we have
\begin{align}
	[G_1^+(E)]^T&=G_1^+(E)\label{G1rels}\\
	G_2^-(-E)~&=-G_2^+(E)\label{G2rels}.
\end{align}
These relations would be useful later on.
The block tri-diagonal structure of $\mathcal{G}(E)$ in Eq.~(\ref{GME}) allows us to  find the required elements of $\mathcal{G}(E)$, which are $\mathcal{G}_{N1}$ and $\mathcal{G}_{11}$, for obtaining the terms in NEGF-expression for conductance. Thus, we  define $I_{2N}$ to be a $2N\cross2N$ identity matrix so that, using the first column of equations from the identity $\mathcal{G}^{-1}(E)\mathcal{G}(E)=I_{2N}$, we can write 
\begin{align}
	&(-E+A-A_\Sigma)\mathcal{G}_{11}+A_W\mathcal{G}_{21}=I_2\\
	&A_W^T \mathcal{G}_{i-1,1}+(-E+A) \mathcal{G}_{i1}+ A_W \mathcal{G}_{i+1,1}=0 ~\text{for}~ 1<i<N\\
	&A_W^T \mathcal{G}_{N-1,1}+(-E+A-A_\Sigma)\mathcal{G}_{N1}=0
\end{align}
We rewrite these equations as a recursion relation, following similar steps as we did for Eq.~(\ref{tmeqt}),  to  obtain 
\begin{equation}
	\begin{pmatrix}
		-I_2\\\mathcal{G}_{11}
	\end{pmatrix}=\Omega_1\Omega^{N-2}_W\Omega_2 \begin{pmatrix}\mathcal{G}_{N1} \\0\end{pmatrix}\label{negftm1}
\end{equation}
with $\Omega_W$ given by Eq.~(\ref{OmegaW}),
\begin{align}
	&\Omega_1=\begin{pmatrix}
		(E-A+A_\Sigma)A_W^{-T} && -A_W\\ A_W^{-T}&&0 
	\end{pmatrix}=  \begin{pmatrix}
		1 && A_\Sigma \\ 0&& 1
	\end{pmatrix} \Omega_{L2}\label{Omega1},\\\label{Omega2}
	&\Omega_2=\begin{pmatrix}
		E-A+A_\Sigma&&1\\1&&0
	\end{pmatrix}=\Omega_{R2}\begin{pmatrix}
		1 && 0 \\ -A_\Sigma && -1
	\end{pmatrix},
\end{align}
where $\Omega_{L2}$ and $\Omega_{R2}$ are the same matrices defined in the scattering calculation by Eq.~(\ref{OmegaL2}) and Eq.~(\ref{OmegaR2}) respectively. Using Eq.~(\ref{Omega1}), Eq.~(\ref{Omega2}) and substituting $A_\Sigma=\frac{\eta_c^2}{\eta_b}e^{iq}I_2$, one can express Eq.~(\ref{negftm1}) as
\begin{equation}
	\begin{pmatrix}
		I_2\\\mathcal{G}_{11}
	\end{pmatrix}=\begin{pmatrix}
		e^{2iq}\mathcal{O} && \bar{\Omega}_{12}+\frac{\eta_b}{\eta_c^2}e^{iq}\bar{\Omega}_{22} \\ \bar{\Omega}_{21}-\frac{\eta_b}{\eta_c^2}e^{iq}\bar{\Omega}_{22}&& -\bar{\Omega}_{22} \end{pmatrix}\begin{pmatrix}\mathcal{G}_{N1} \\0\end{pmatrix},\label{negftm}
\end{equation}
where $\mathcal{O}$ is given by Eq.~(\ref{McalO}). From the upper block of Eq.~(\ref{negftm}), we obtain the following matrix equation for $[G^+(E)]_{N1}$ and $[G_2^+(-E)]_{N1}^*$:
\begin{align}
	&\ket{+}=-e^{2iq}\mathcal{O}\left[[G^+_1(E)]_{N1}\ket{+}+[G_2^+(-E)]_{N1}^*\ket{-}\right].
	%&\ket{-}=e^{2iq}\mathcal{O}\left[[G^+_2(E)]_{N1}\ket{+}+[G_1^+(-E)]_{N1}^*\ket{-}\right],
\end{align}
which gives two linear  equations for $[G^+(E)]_{N1}$ and $[G_2^+(-E)]_{N1}^*$:
\begin{align}
	&1=- e^{2iq}[G^+_1(E)]_{N1} \bra{+}\mathcal{O}\ket{+}-e^{2iq}[G_2^+(-E)]_{N1}^* \bra{+}\mathcal{O}\ket{-}\\
	&0= -e^{2iq}[G^+_1(E)]_{N1} \bra{-}\mathcal{O}\ket{+}-e^{2iq}[G_2^+(-E)]_{N1}^* \bra{-}\mathcal{O}\ket{-}.
	%&1= e^{2iq}[G^+_2(E)]_{N1} \bra{-}\mathcal{O}\ket{+}+e^{2iq}[G_1^+(-E)]_{N1}^* \bra{-}\mathcal{O}\ket{-}\\
	%&0= e^{2iq}[G^+_2(E)]_{N1} \bra{+}\mathcal{O}\ket{+}+e^{2iq}[G_1^+(-E)]_{N1}^* \bra{+}\mathcal{O}\ket{-}
\end{align}
Comparing these with Eq.~(\ref{tntascateq1} and \ref{tntascateq2}) and noticing that $\sin q= \eta_b g(E)$ we have
\begin{align}
	2i\eta_c^2g(E)[G^+_1(E)]_{N1}&=e^{-2iq}t_n, \label{amptnT1}\\
	2i\eta_c^2g(E)[G_2^+(-E)]_{N1}^*&=-e^{-2iq}t_a. \label{amptaT2}
\end{align}
These equations, along with Eq.~(\ref{G1rels} and \ref{G2rels}) imply that
\begin{align}
	T_1(E)=4\eta_c^4g^2(E)|[G_1^+(E)]_{1N}|^2&=|t_n|^2,\label{tnT1}\\
	T_2(E)=4\eta_c^4g^2(E)|[G_2^+(E)]_{1N}|^2&=|t_a|^2, \label{taT2}
\end{align}
which are the required relations. If we consider the lower block equations of Eq.~(\ref{negftm}), then one of the component equation reads

\begin{align}
	&[G_2^+(-E)]^*_{11}=\left[\bra{-}\bar{\Omega}_{21}\ket{+}-e^{iq}\frac{\eta_c^2}{\eta_b}\bra{-}\bar{\Omega}_{22}\ket{+}\right][G_1^+(E)]_{N1}\notag\\&+\left[\bra{-}\bar{\Omega}_{21}\ket{-}-e^{iq}\frac{\eta_c^2}{\eta_b}\bra{-}\bar{\Omega}_{22}\ket{-}\right][G_2^+(-E)]^*_{N1}. 
\end{align}
We replace $[G_1^+(E)]_{N1}$ and $[G_2^+(-E)]^*_{N1}$ for $t_n$ and $t_a$ in this equation with the help of  Eq.~(\ref{amptnT1}, \ref{amptaT2}). Comparing the resulting equation with Eq.~(\ref{raexp}) and using Eq.~(\ref{G2rels}), we finally obtain
\begin{equation}
	T_3(E)=T_3(-E)=4\eta_c^4g^2(E)|[G_2^+(E)]_{11}|^2=|r_a|^2.\label{raT3}
\end{equation}
This completes the  analytic proof for the equivalence.  

We now proceed to the next section where we present numerical comparison of the results from the two approaches  and  the  behaviour of the zero bias conductance peak in different parameter regimes of the Hamiltonian, along with some other numerical results which are useful in discussing the electron transport in the wire.
 
\section{Numerical Results and Discussion}
\label{numerical}
\begin{figure}[htb!]
 	\centering
 	\subfigure[$~N=50$]{
 		\includegraphics[width=4.cm,height=31mm]{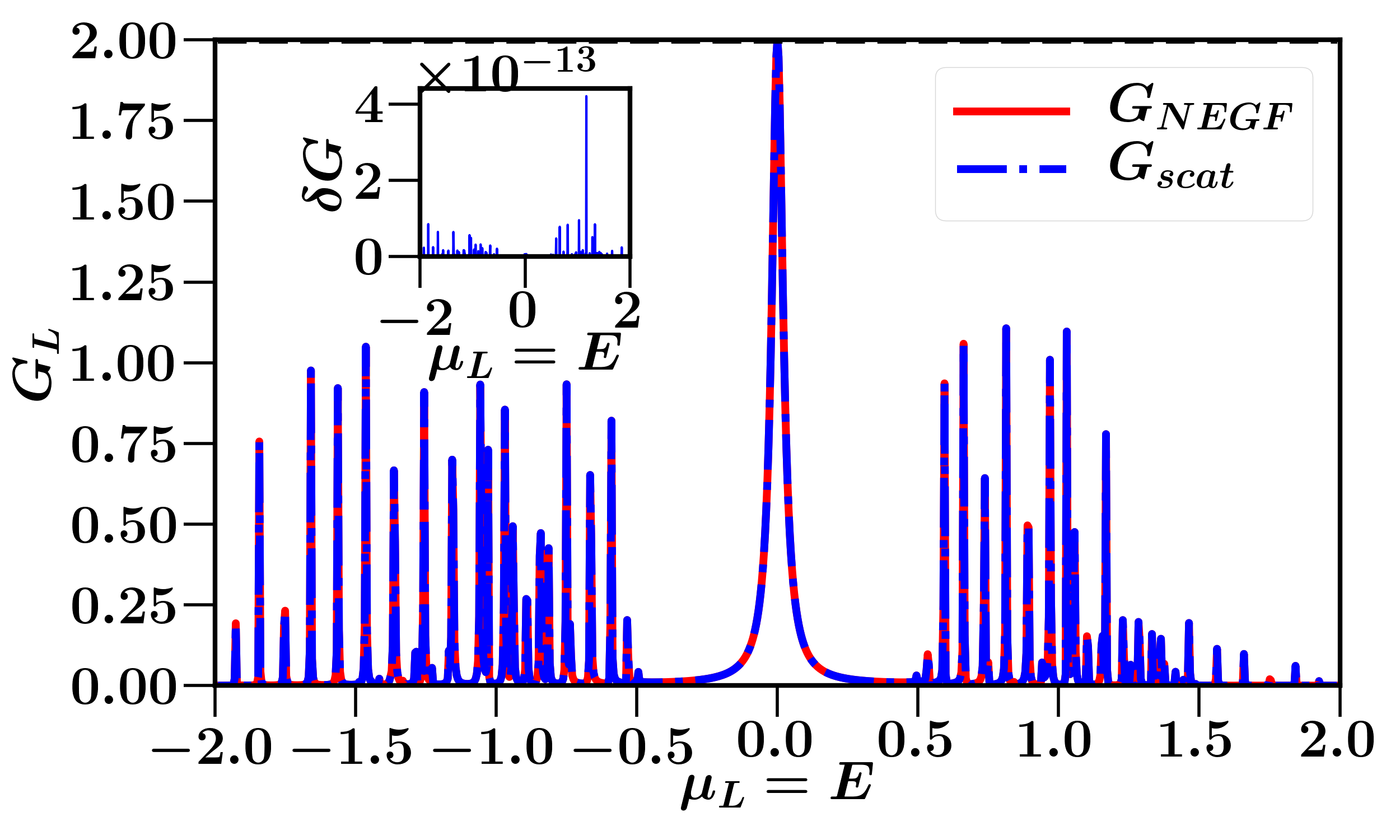}
 	}
 	\subfigure[$~N=3$]{
 		\includegraphics[width=4.cm,height=31mm]{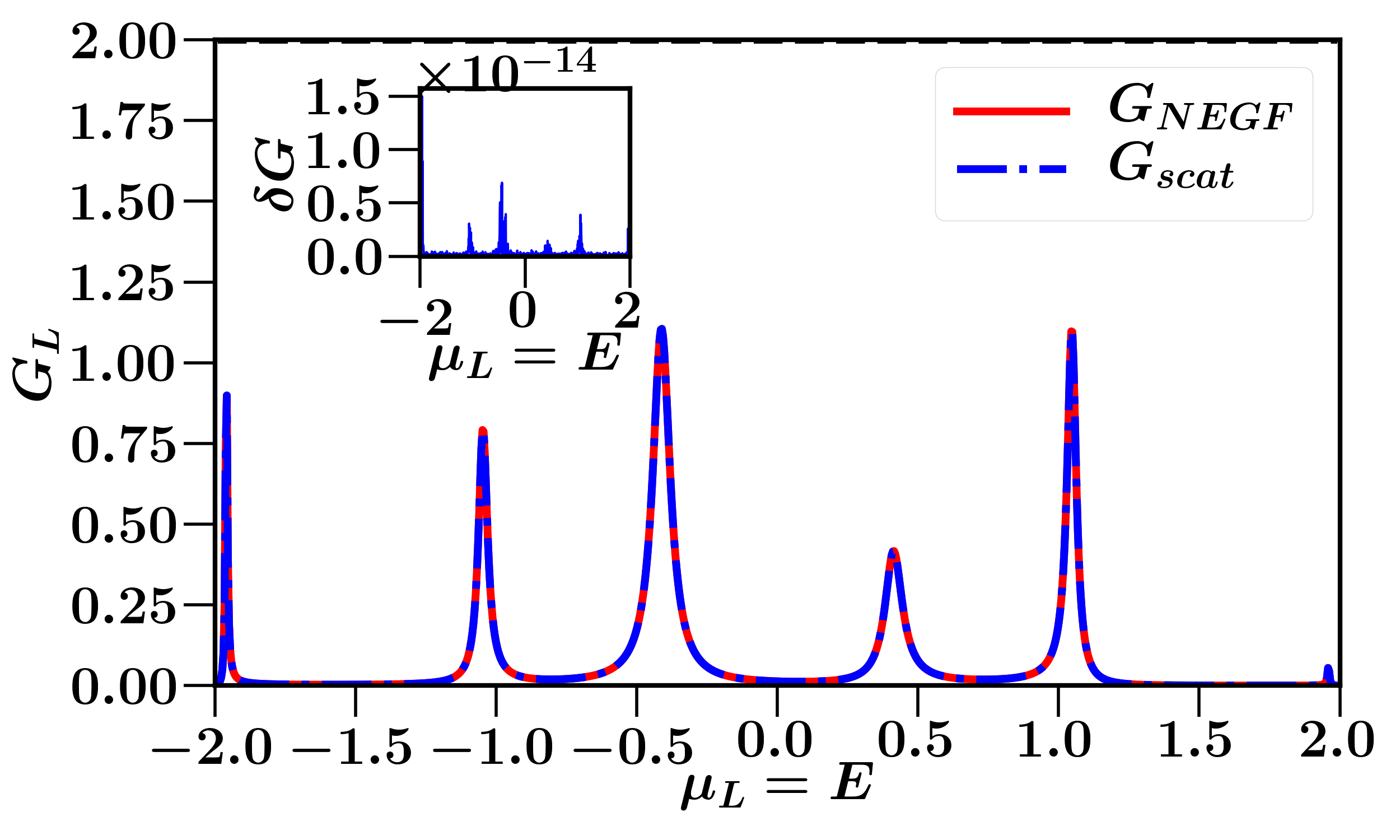}
 	}
 	\subfigure[$~N=50$]{
 		\includegraphics[width=4.cm,height=31mm]{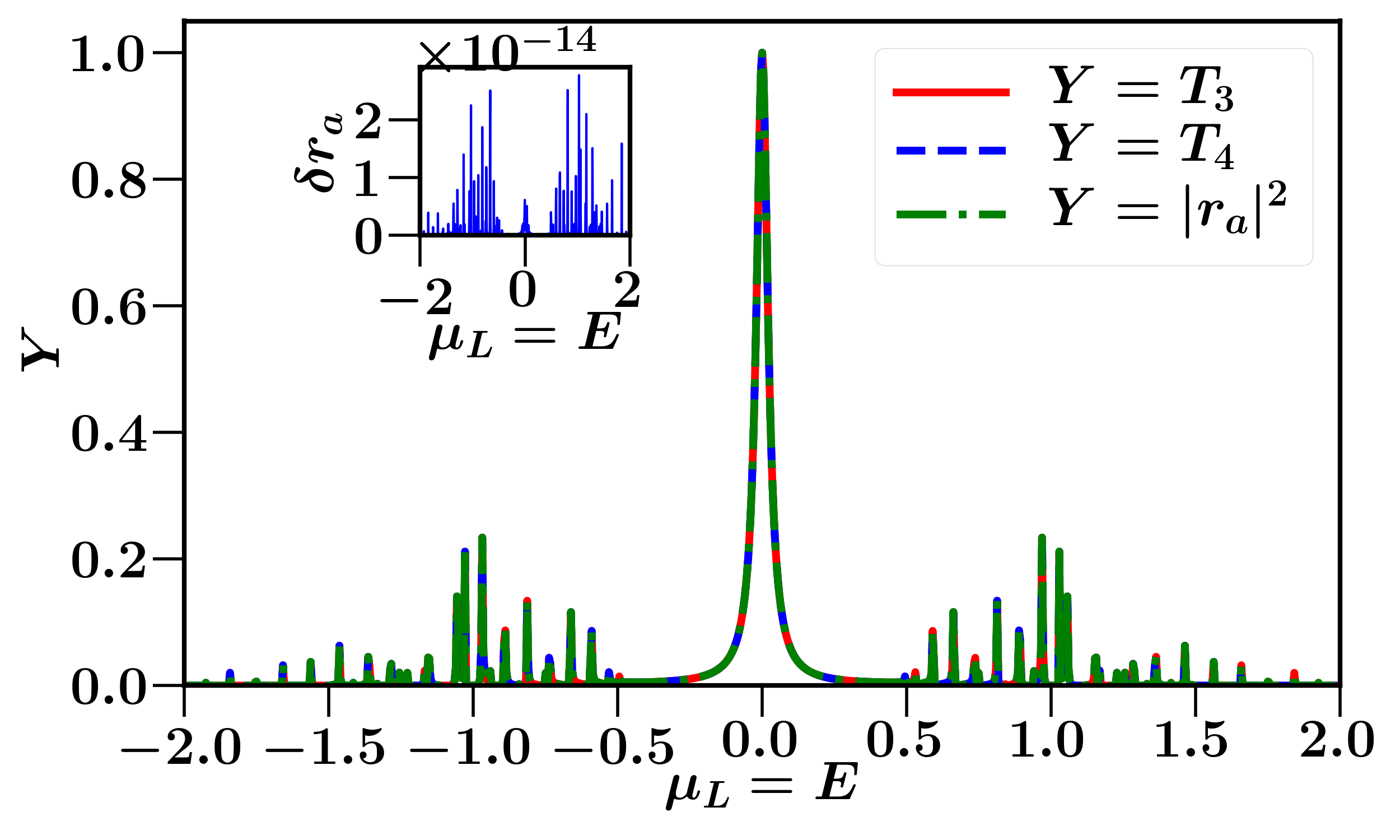}
 	}
 	\subfigure[$~N=3$]{
 		\includegraphics[width=4.cm,height=31mm]{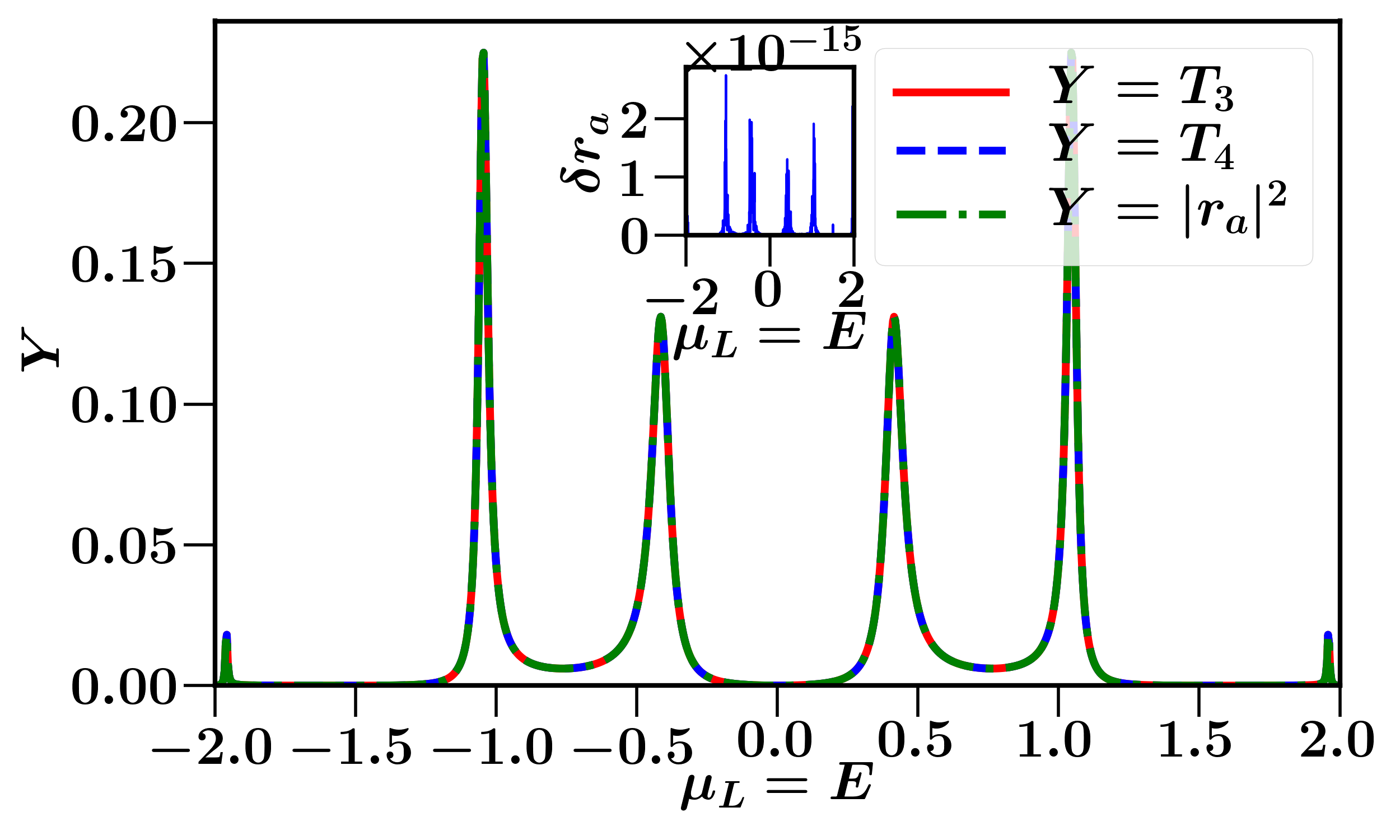}
 	}
 	\subfigure[$~N=50$]{
 		\includegraphics[width=4.cm,height=31mm]{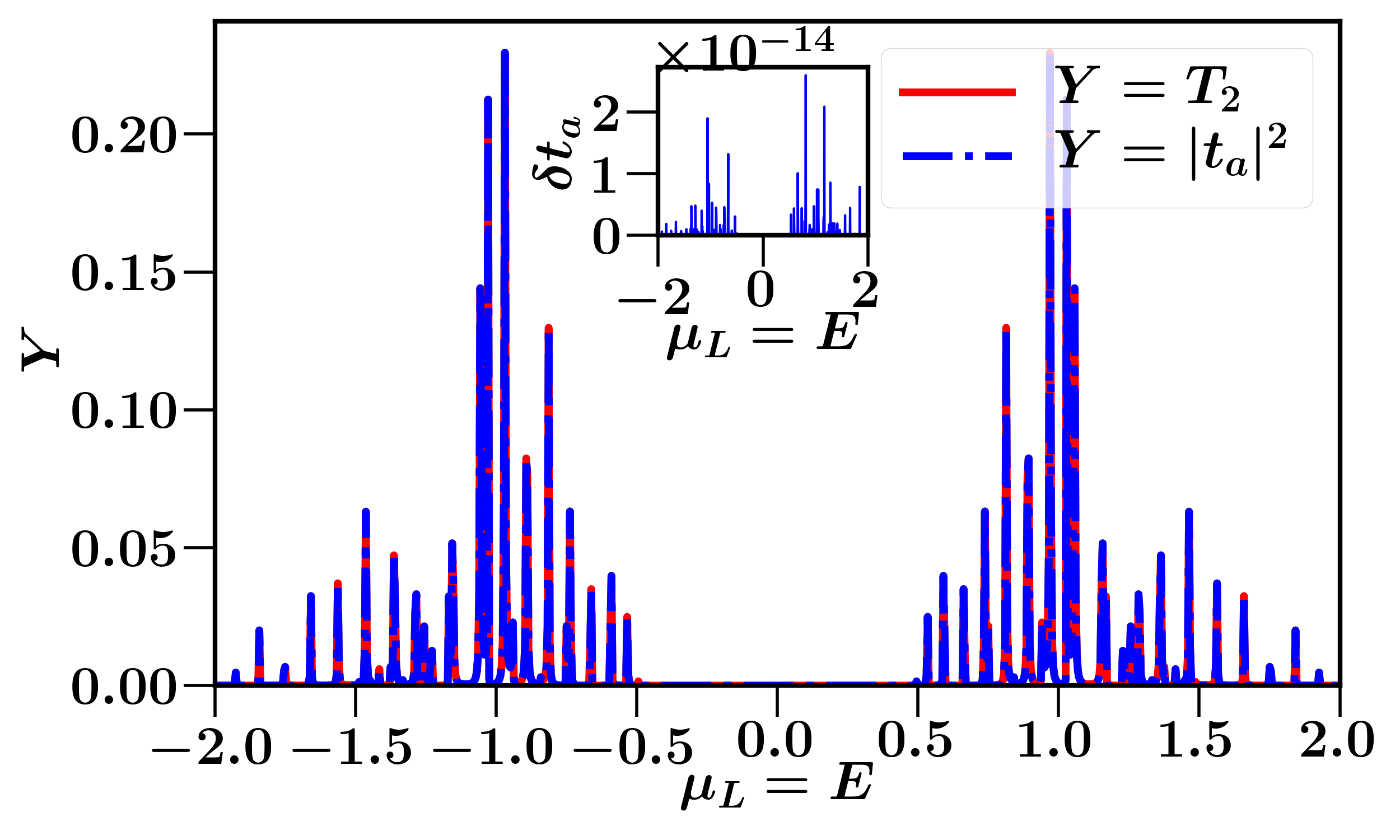}
 	}
 	\subfigure[$~N=3$]{
 		\includegraphics[width=4.cm,height=31mm]{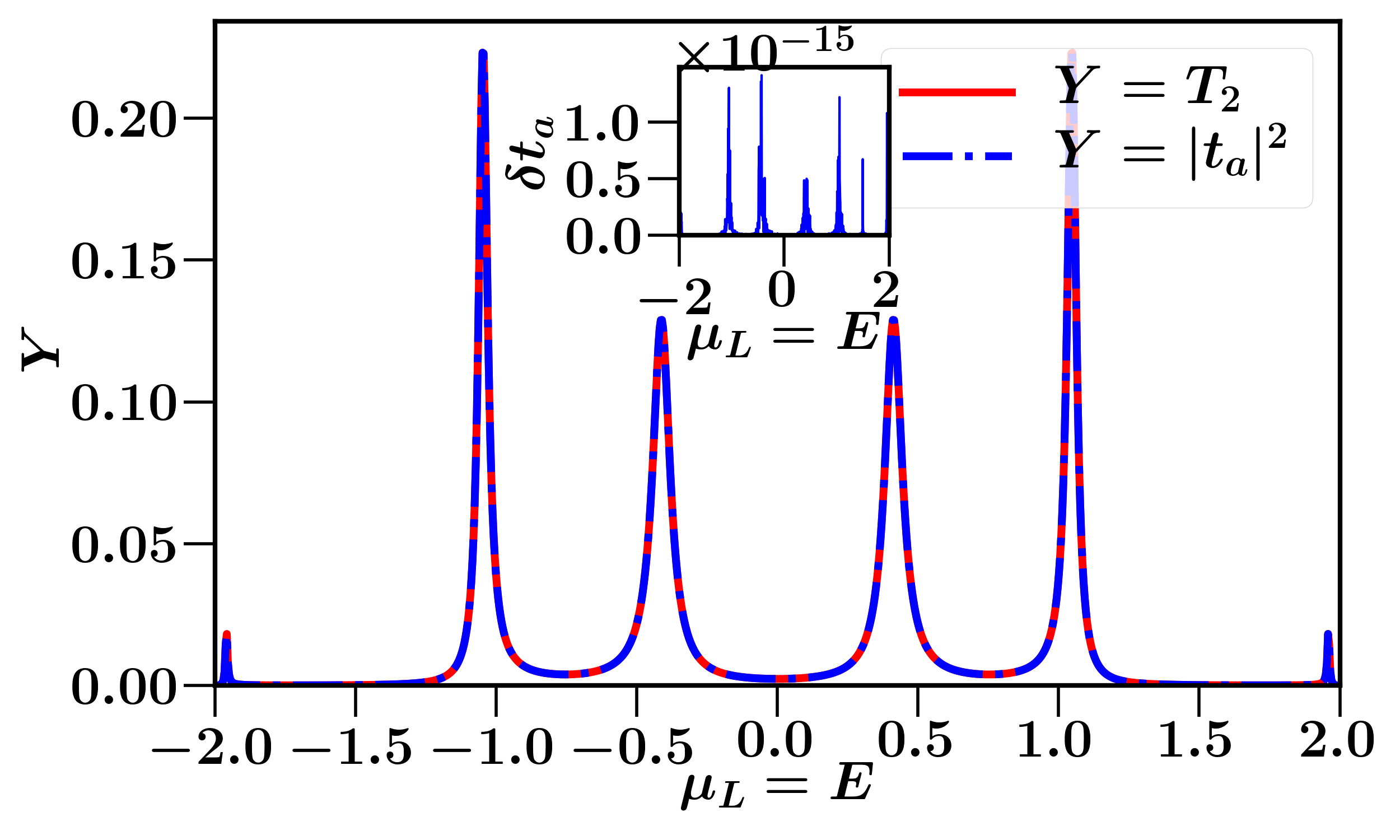}
 	}
 	\subfigure[$~N=50$]{
 		\includegraphics[width=4.cm,height=31mm]{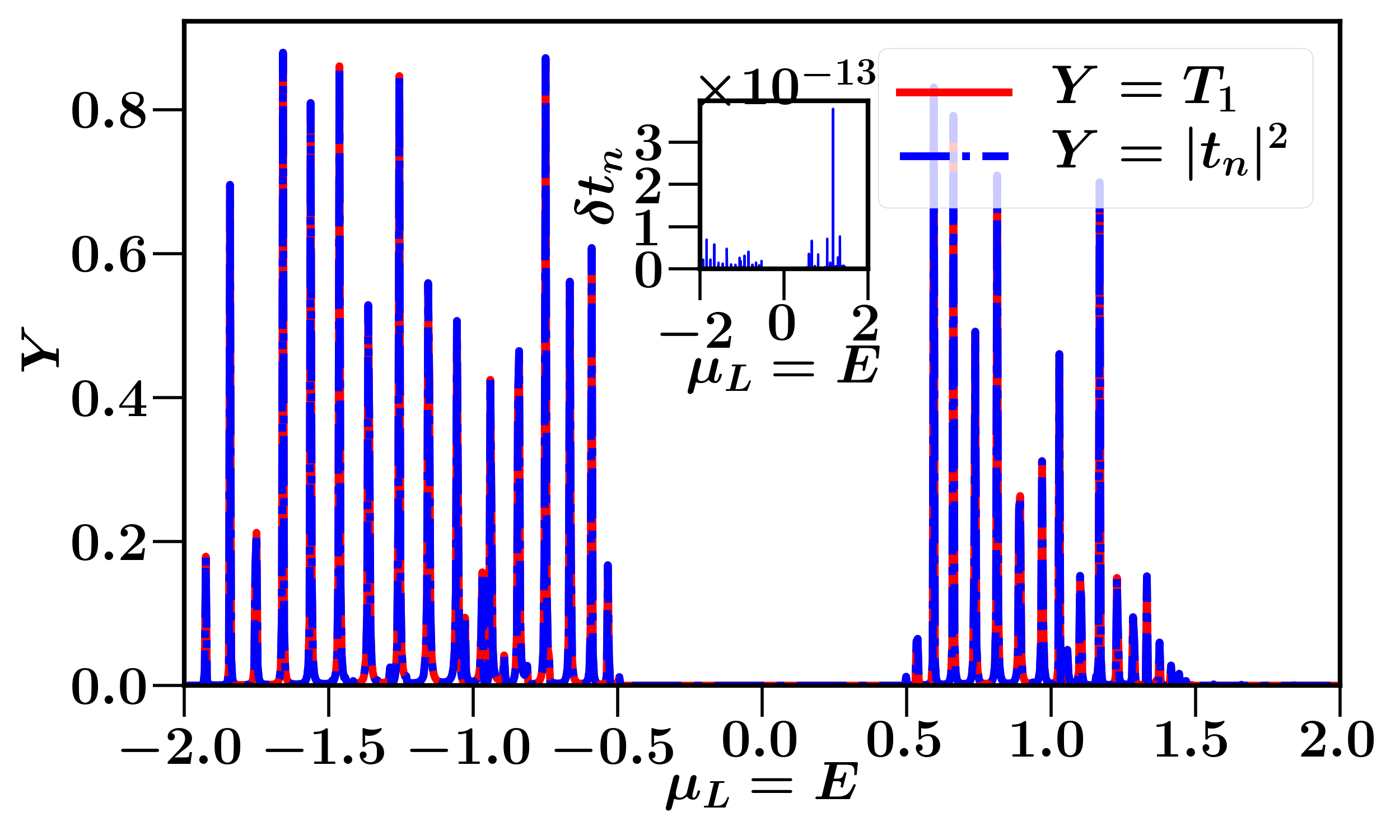}
 	}
 	\subfigure[$~N=3$]{
 		\includegraphics[width=4.cm,height=31mm]{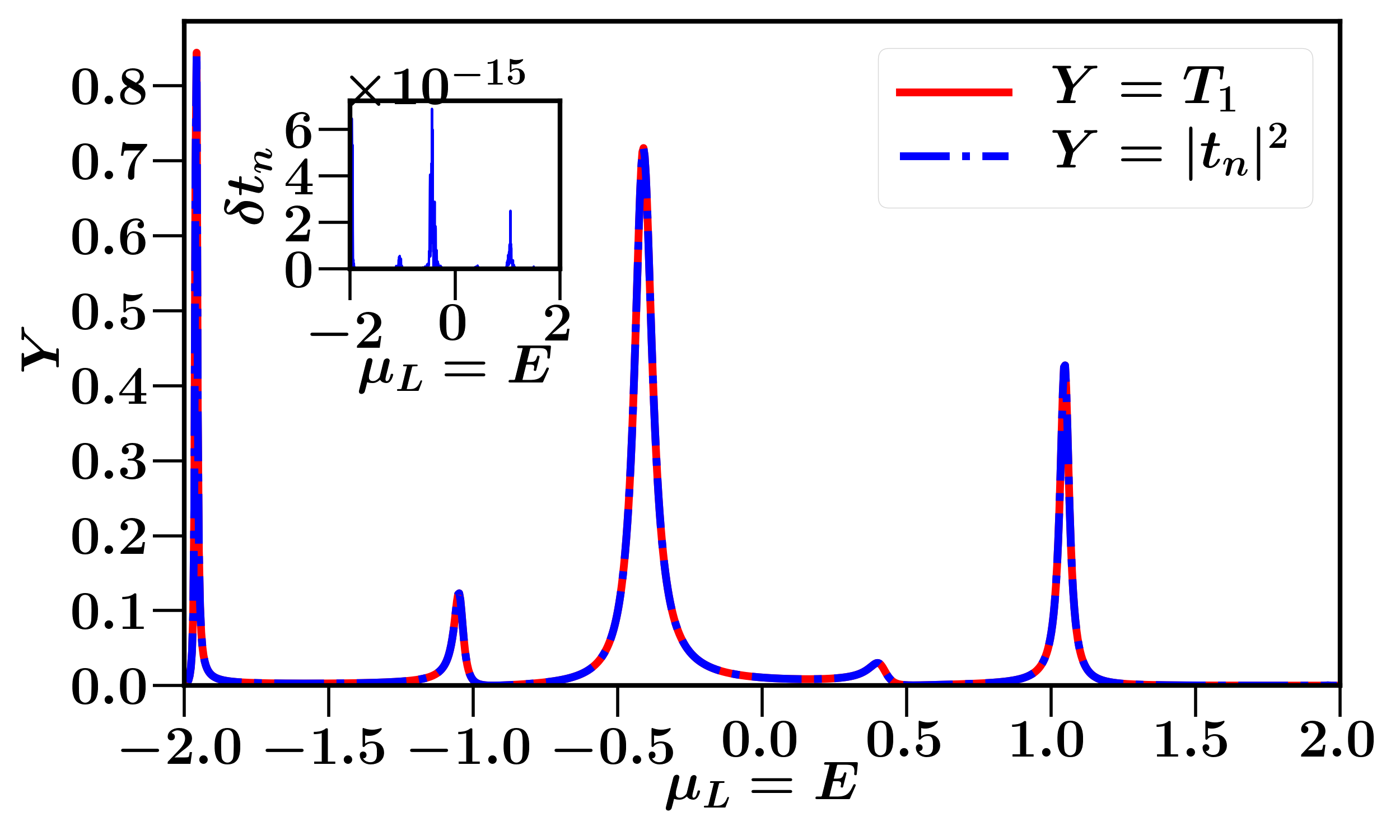}
 	}
 	\caption{The comparison of various quantities obtained from QLE-NEGF and scattering approaches for parameter values --- $\Delta=0.25$, $\eta_c=0.2$
 		$\eta_b=1$, $\eta_w=1$ and  $\mu_w=0.5$. The inset in each plot shows the absolute value of the difference between the corresponding curves. (a) shows the comparison of NEGF-expression and the scattering expression for $G_L$. From the  corresponding inset, which shows the difference between the two results, we can say that the two expressions match perfectly. (c) shows the  comparison of $\abs{r_a}^2$ with the last two terms, $T_3=T_3(\mu_L)$ and $T_4=T_3(-\mu_L)$ in the NEGF-expression for conductance. In the inset, $\delta r_a=\abs{T_3-\abs{r_a}^2}+\abs{T_4-\abs{r_a}^2}$ is plotted with $\mu_L=E$. (e) and (g) show the comparison of $\abs{t_n}^2$ and $\abs{t_a}^2$ with $T_1=T_1(\mu_L)$ and $T_2=T_2(\mu_L)$ of the NEGF-conductance expression respectively. The corresponding insets of the two plots show the variation of $\delta t_n=|T_1-\abs{t_n}^2|$ and  $\delta t_a=|T_2-\abs{t_a}^2|$ respectively. Plots (b), (d), (f) and (h) show same results as (a), (c), (e) and (g) respectively for $N=3$.}
 	\label{plts}
 \end{figure}

The quantities required for calculating the conductance from the scattering method are simply obtained by solving the system of eight linear equations given by Eq.~(\ref{ampeqns})  while the terms in the NEGF expression for conductance are straight forward to calculate via Eq.~(\ref{st1}-\ref{st3}).   We note that $\mu_L$ in the NEGF expression for conductance plays the role of $E$ in the scattering method.  In Figs.~(\ref{plts}a) and (\ref{plts}b) we show the conductance from the NEGF-expression and the scattering expression plotted as functions of $\mu_L=E$ for a long chain ($N=50$) and a short chain ($N=3$) respectively. The plot shows a perfect agreement between the two. Similarly,  Figs.~(\ref{plts}c,\ref{plts}e,\ref{plts}g) for $N=50$, and  Figs.~(\ref{plts}d,\ref{plts}f,\ref{plts}h) for $N=3$, show a perfect agreement between the quantities $T_3(\mu_L)$, $T_3(-\mu_L)$ and $\abs{r_a}^2$, $T_2(\mu_L)$ and $\abs{t_a}^2$,  $T_1(\mu_L)$ and $\abs{t_n}^2$ respectively. 

It is interesting to consider the spectrum and the form of wavefunctions of the Kitaev chain connected to leads. With infinite leads we saw in Sec.~\ref{scattering} that all scattering eigenstates and the MBS can be obtained analytically. We can also work with large finite reservoirs and obtain the spectrum by diagonalising the matrix $\mathcal{A}$ in Eq.~\eqref{Amatrix}.  Note that this furnishes a doubly counted spectrum of the Hamiltonian and the corresponding wavefunctions have twice the number of components as the number of sites in the system. As per the basis that we have chosen to write $\mathcal{A}$, two adjacent components will correspond to a single site of the system. Therefore in our plots, if $N_L=N_R=N_B$  then left bath sites are  from $-2N_B+1$ to $0$, the wire sites are from $1$ to $2N$ and the sites from $2N+1$ to $2N+2N_B$  correspond to the right reservoir.  The wavefunctions were normalized to one and we typically uses a reservoir size of $N_B=1000$.

\begin{figure}[htb!]
	\centering
	\subfigure[]{
		\includegraphics[width=4.cm,height=35mm]{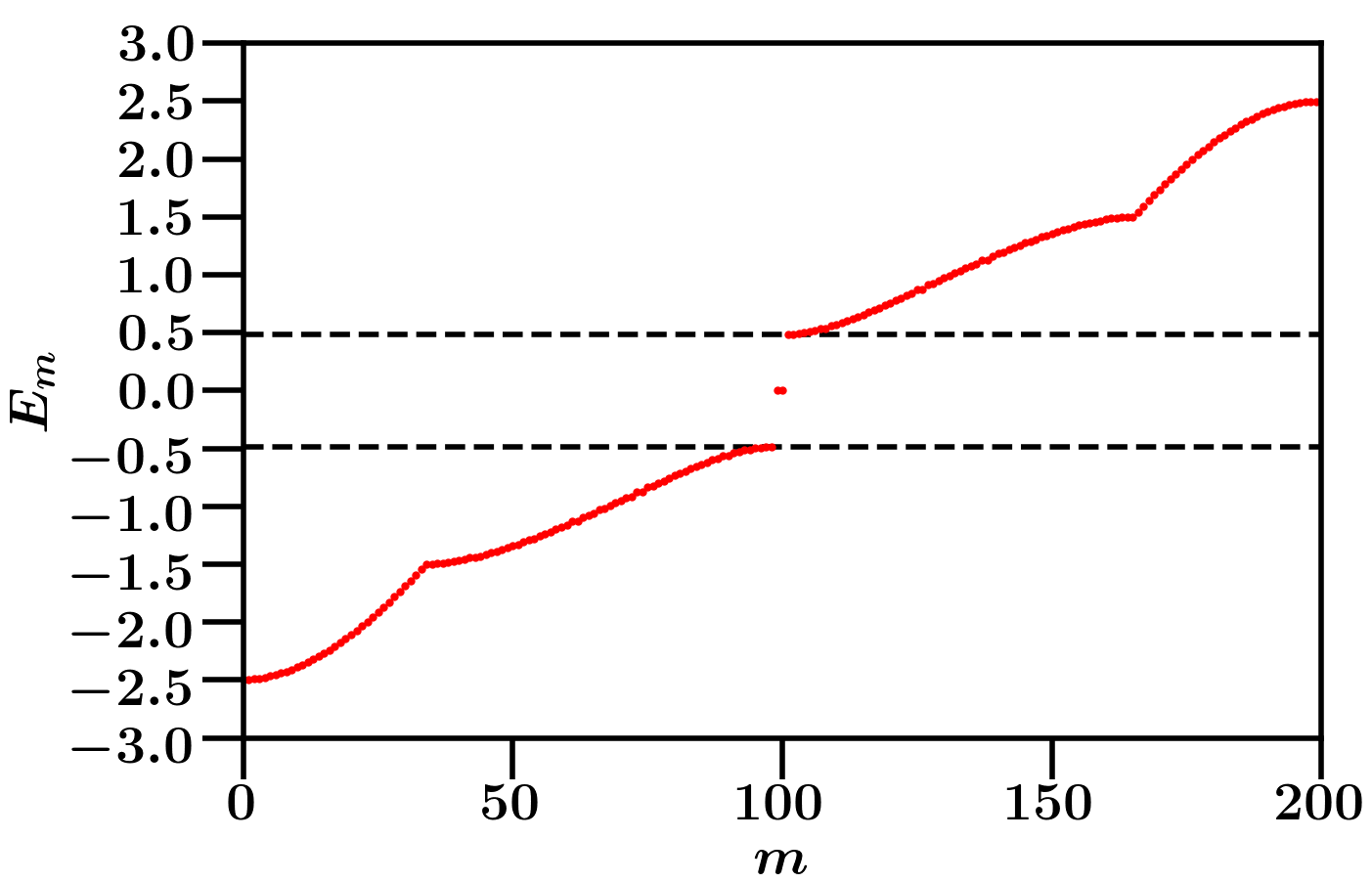}
	}
	\subfigure[]{
		\includegraphics[width=4.cm,height=35mm]{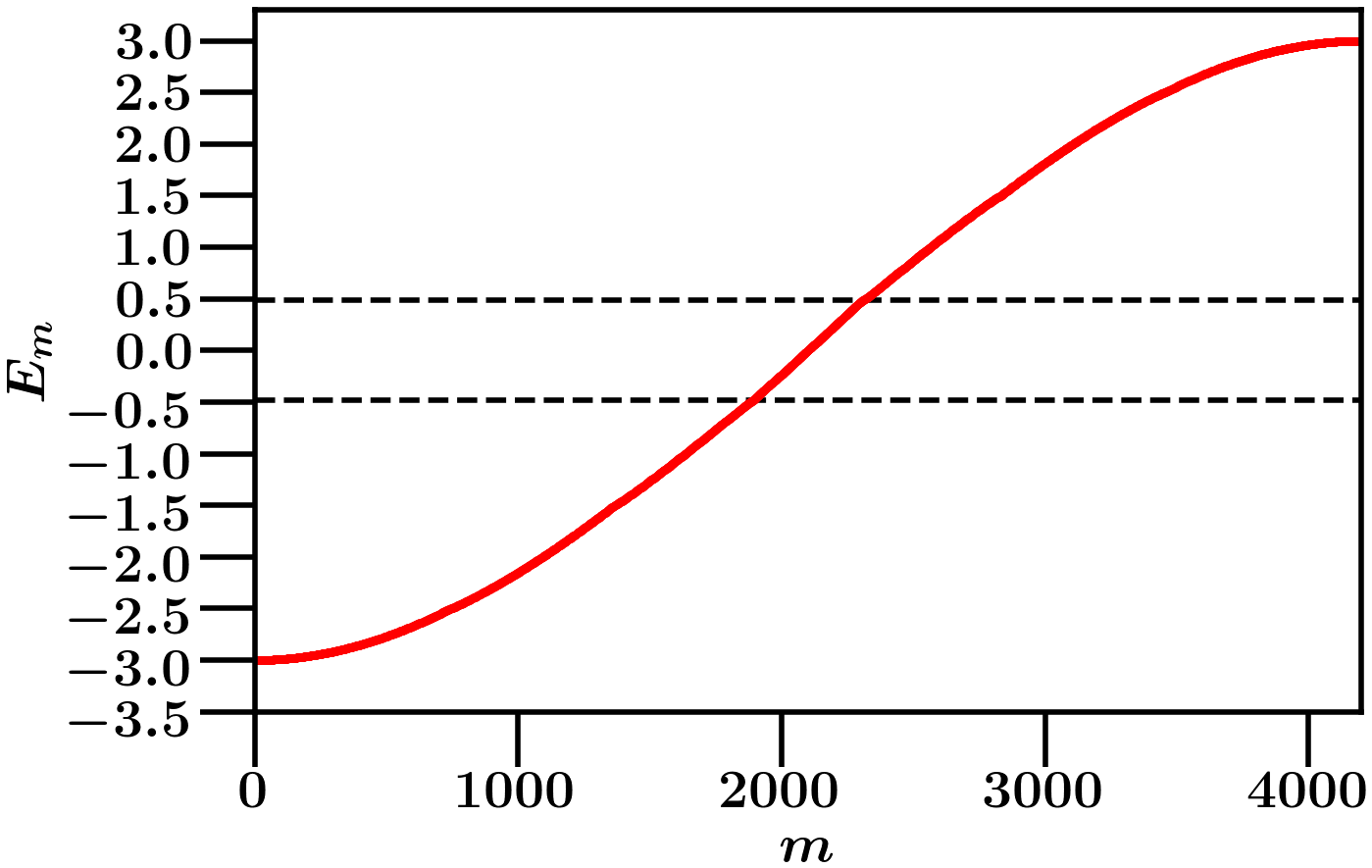}
	}
	\subfigure[]{
		\includegraphics[width=4.cm,height=35mm]{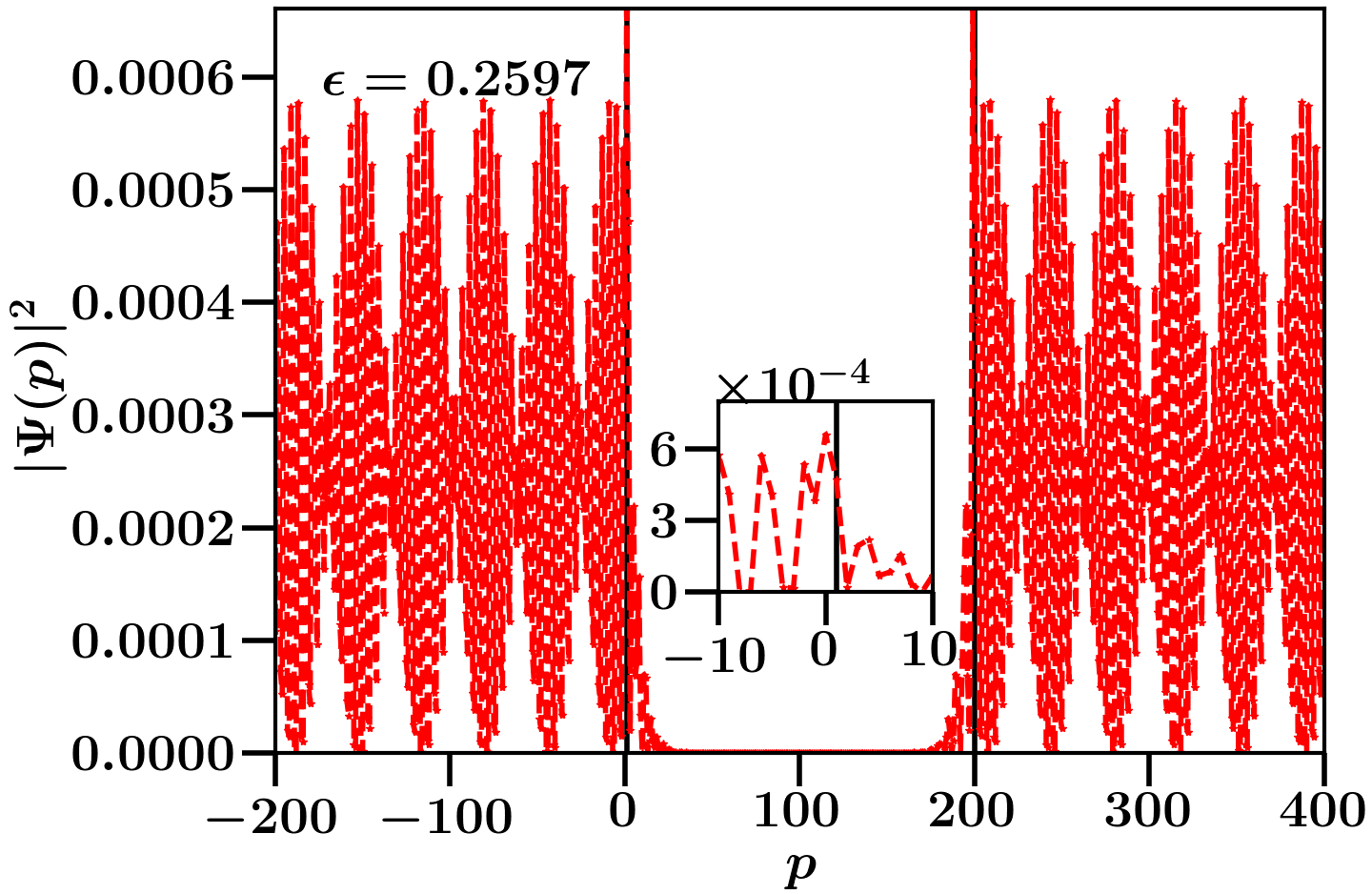}
	}
	\subfigure[]{
		\includegraphics[width=4.cm,height=35mm]{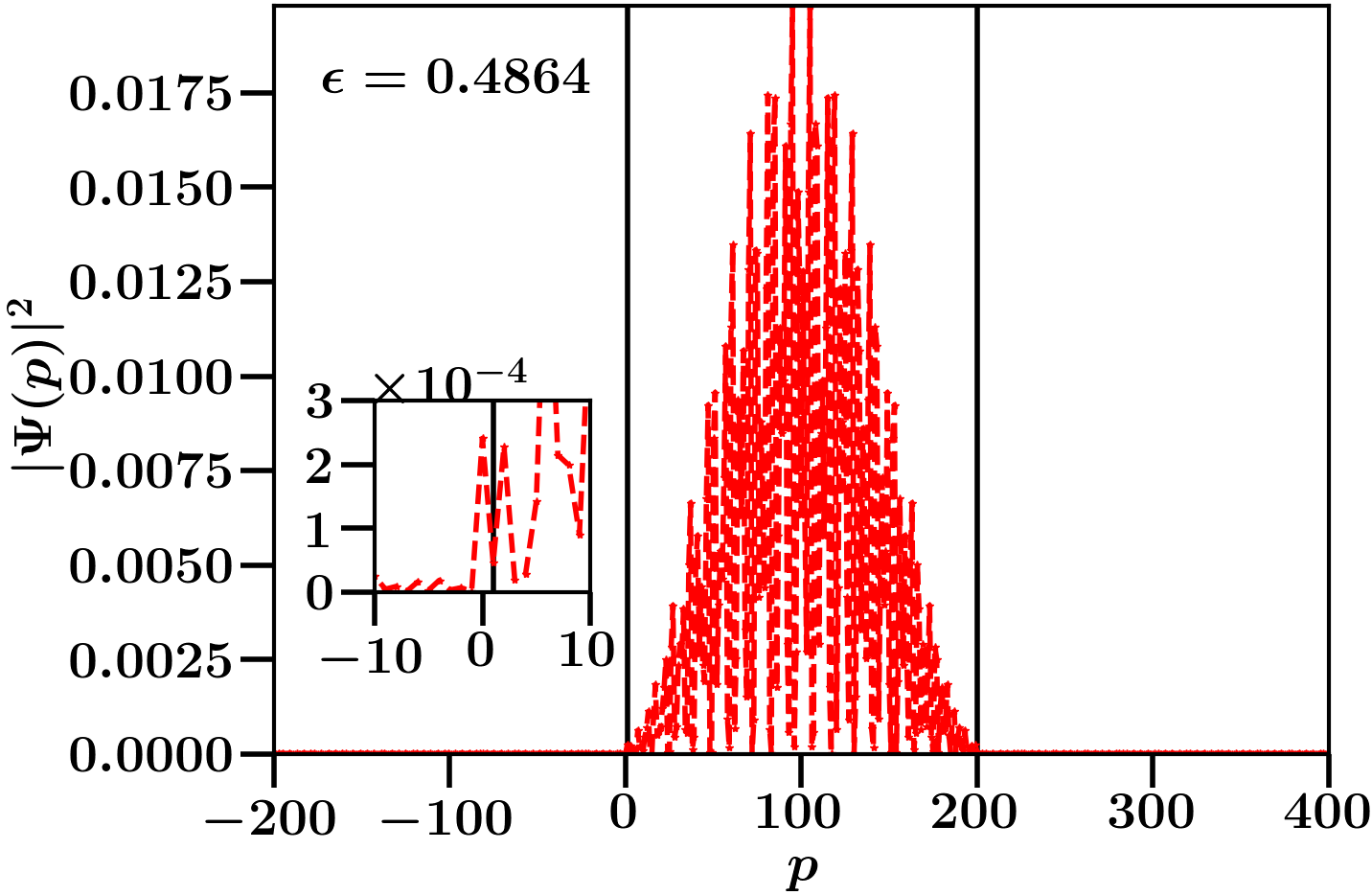}
	}
	\caption{(a) shows the spectrum of a wire  in absence of reservoirs(isolated wire) at parameter values--$N=100$, $\eta_w=1$, $\Delta=0.25$ and $\mu_w=0.5$. (b) shows the spectrum of the wire connected to the two reservoirs with parameter values--$N_B=1000$, $N=100$, $\eta_w=1$, $\Delta=0.25$ , $\mu_w=0.5$, $\eta_b=1.5$ and $\eta_c=0.5$. The two horizontal lines are at $E\approx\pm 0.486$ in these plots mark the SC gap at these parameter values. The two points between the SC gap in (a) are the zero energy MBS of the isolated wire. The spectrum in (b) also has similar  zero modes but cannot be distinctly seen. (c) and (d) show two eigenstates of the matrix $\mathcal{A}$ in Eq.~(\ref{H2}) which serve as the effective eigenstates of the joint system. In these two plots, the two vertical lines at $p=1$ and $p=200$  mark the two ends of the wire and the insets show the left junction zoomed in.}
	\label{specplts}
\end{figure}

\begin{figure}[htb!]
	\centering
	\subfigure[]{
		\includegraphics[width=4.cm,height=35mm]{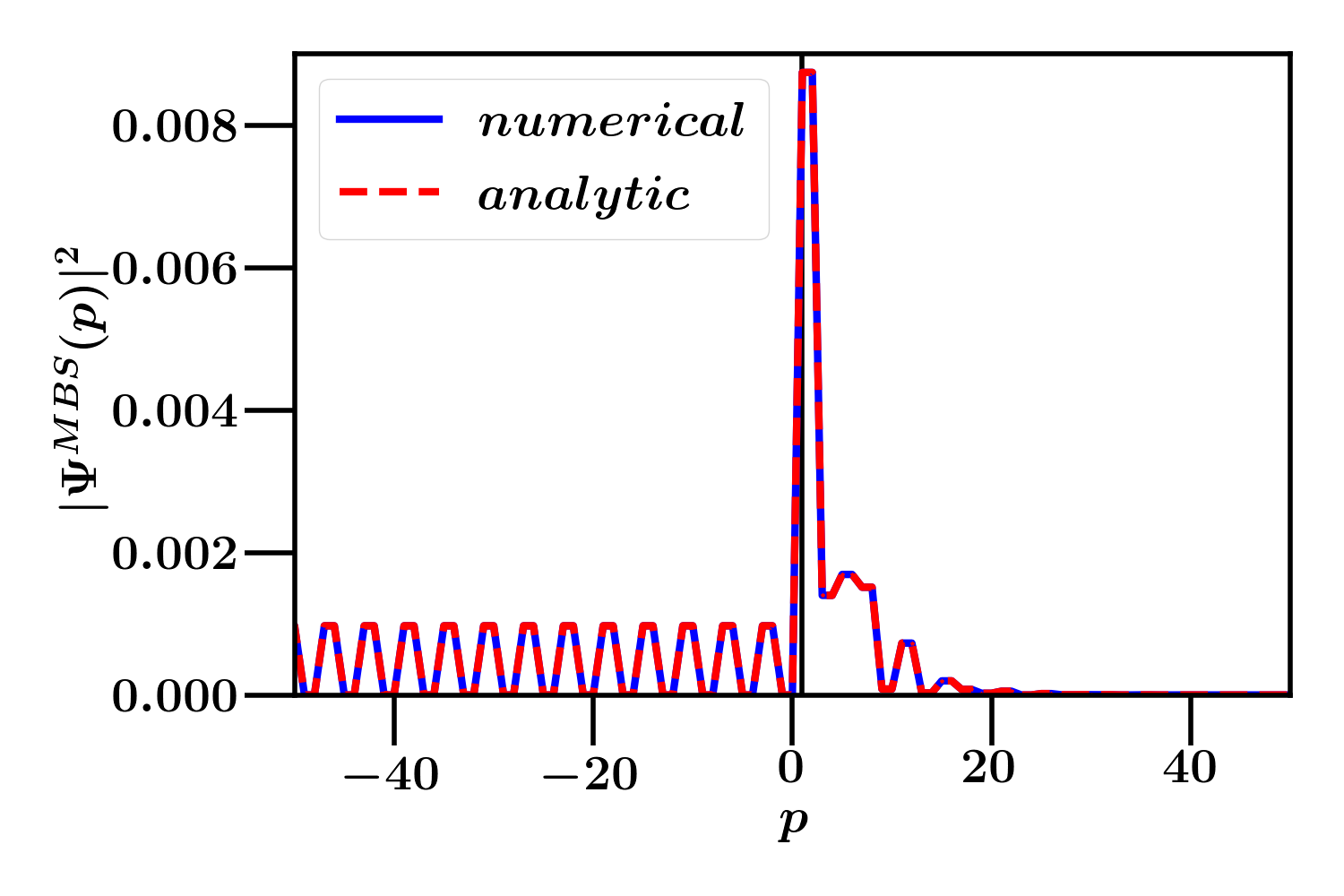}
	}
	\subfigure[]{
		\includegraphics[width=4.cm,height=35mm]{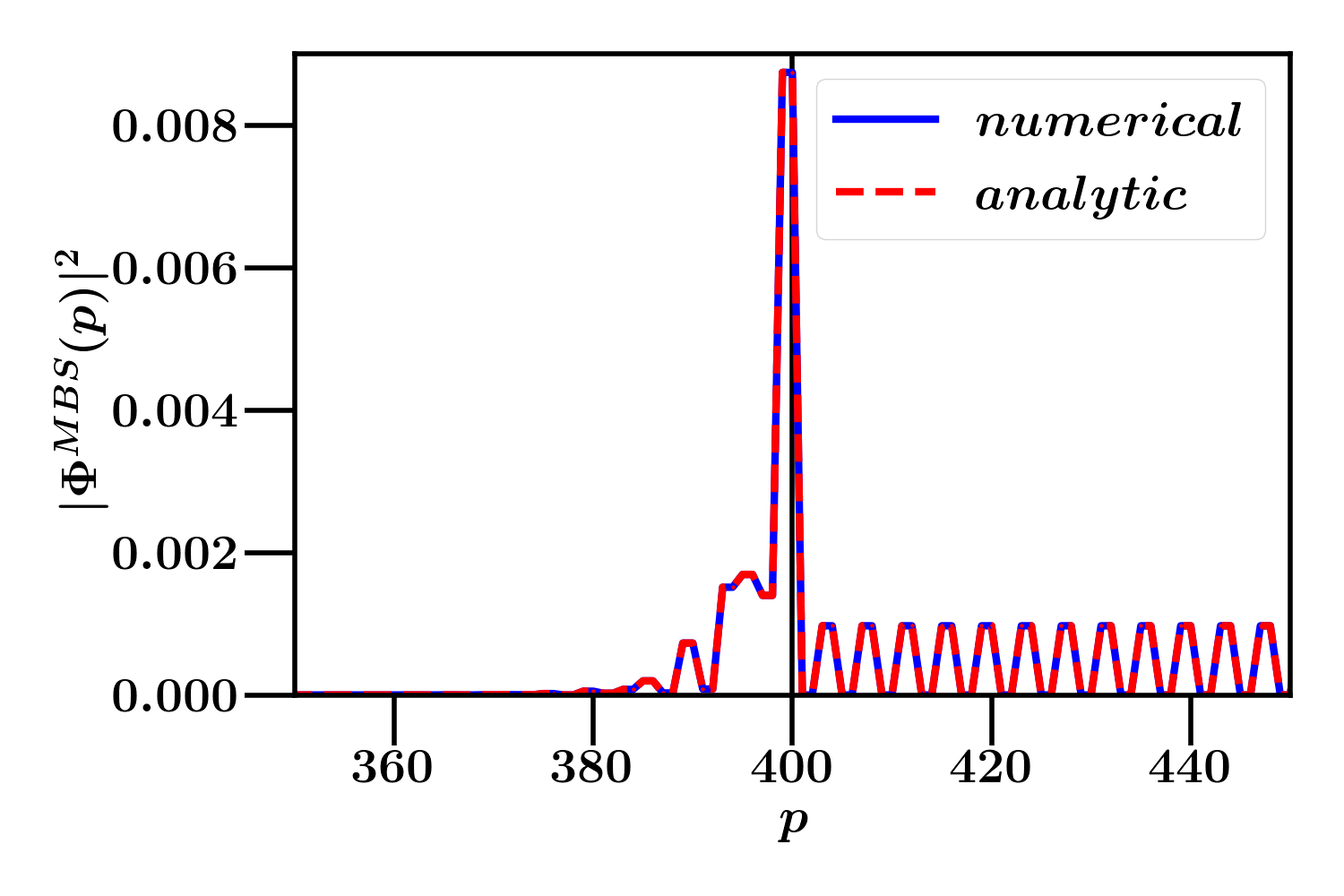}
	}
	\subfigure[]{
		\includegraphics[width=4.cm,height=35mm]{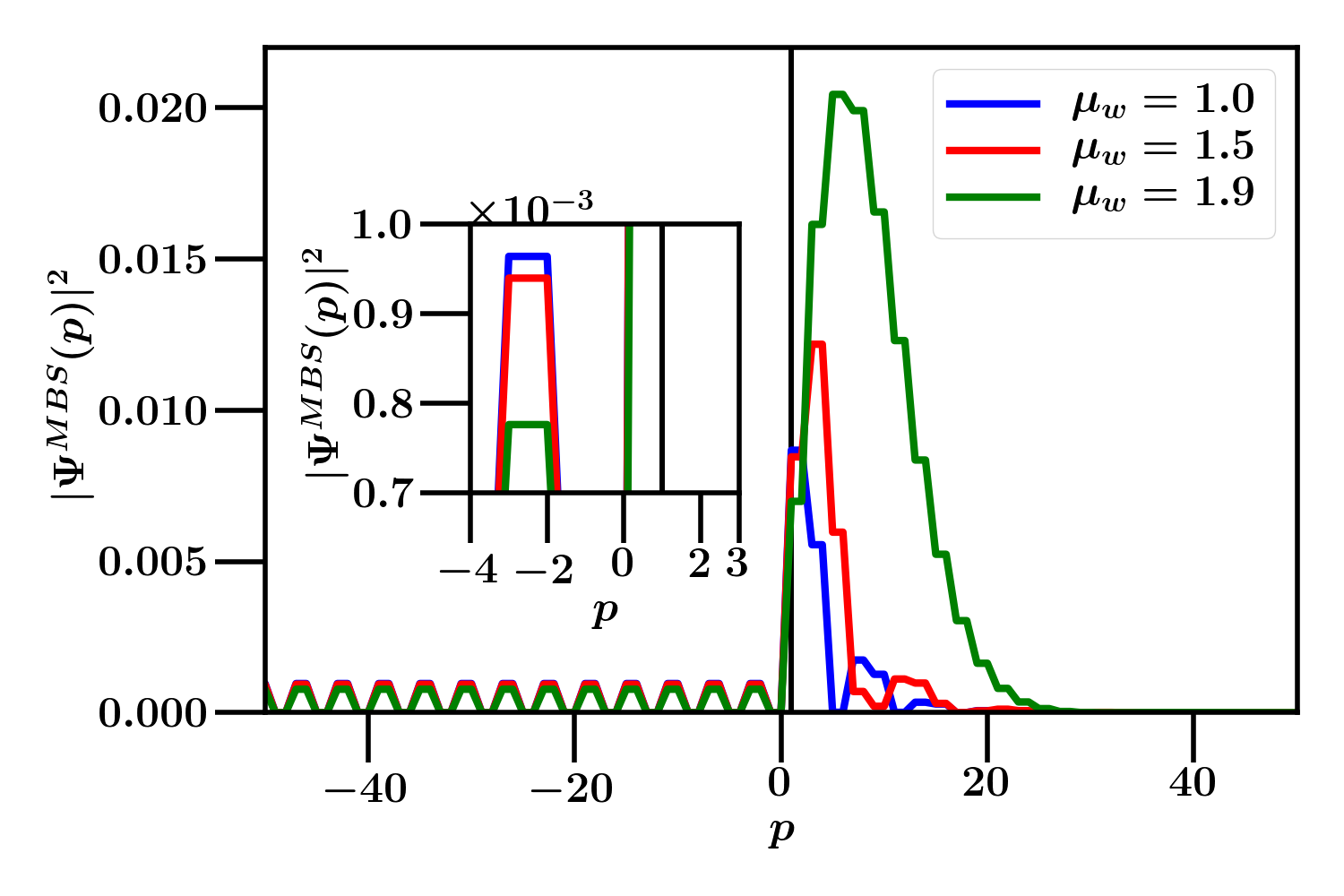}
	}
	\subfigure[]{
		\includegraphics[width=4.cm,height=35mm]{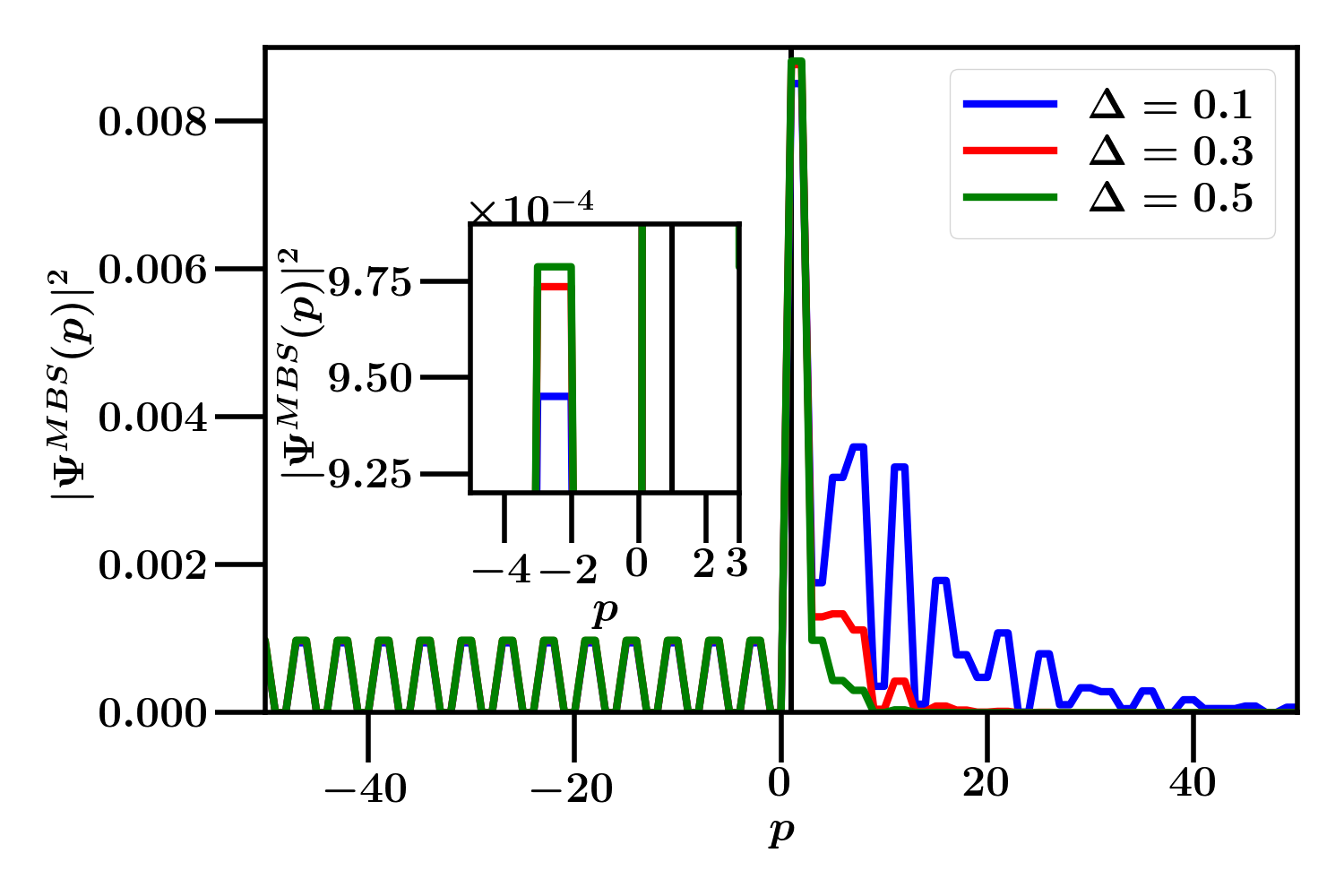}
	}
	\caption{Plots (a) and (b) show the two MBS which are localized at the left end and the right end of the wire respectively with parameter values-- $\eta_b=1.5$, $\mu_w=\eta_c=0.5$, $\Delta=0.25$, $\eta_w=1$ and $N=200$. In these two plots, we choose $N_B=1000$ for numerical calculation of the MBS and we take the same size of the reservoirs to normalize the analytical wavefunctions for the two MBS. (c) and (d) show the analytical MBS wavefunction for different $\mu_w$ and $\Delta$ with the other parameters same as in (c) and (d) respectively.  The insets show the wavefunction in the reservoirs zoomed in.} %The insets show the wavefunction inside the reservoirs zoomed in and it can be clearly seen from the two inset plots that the weight of the MBS in the reservoirs increases as we decrease $\mu_w$ or increase $\Delta$. 
	\label{mbsplts}
\end{figure}

\begin{figure}[htb!]
 	\centering
 	\subfigure[ Zero bias Peak at different $\eta_c$ with $\eta_b=1.5$, $\mu_w=0.5$ and $\Delta=0.1$]{
 		\includegraphics[width=4.cm,height=33mm]{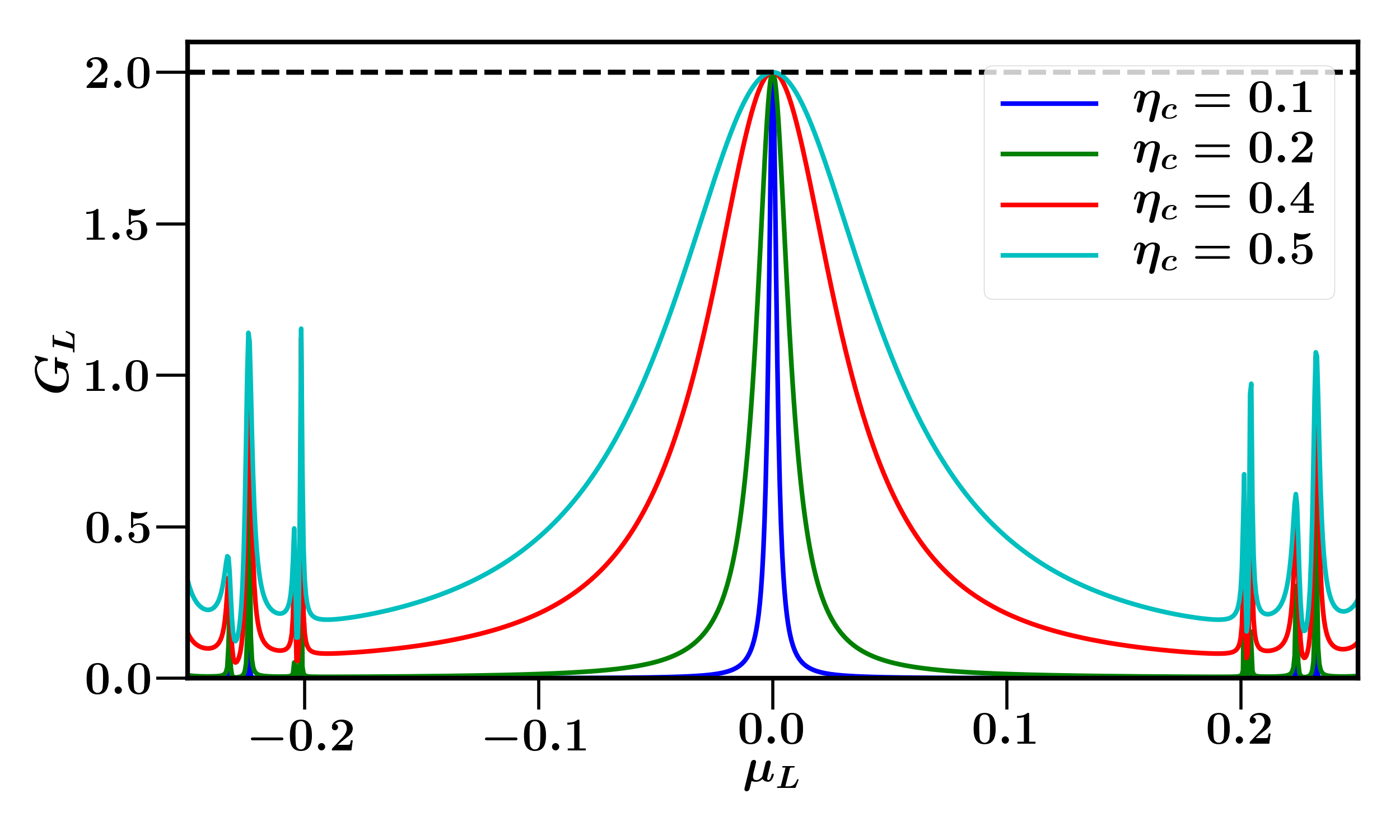}
 	}
 	\subfigure[ Zero bias Peak at different $\eta_b$ with $\mu_w=0.5$, $\eta_c=0.3$ and $\Delta=0.1$]{
 		\includegraphics[width=4.cm,height=33mm]{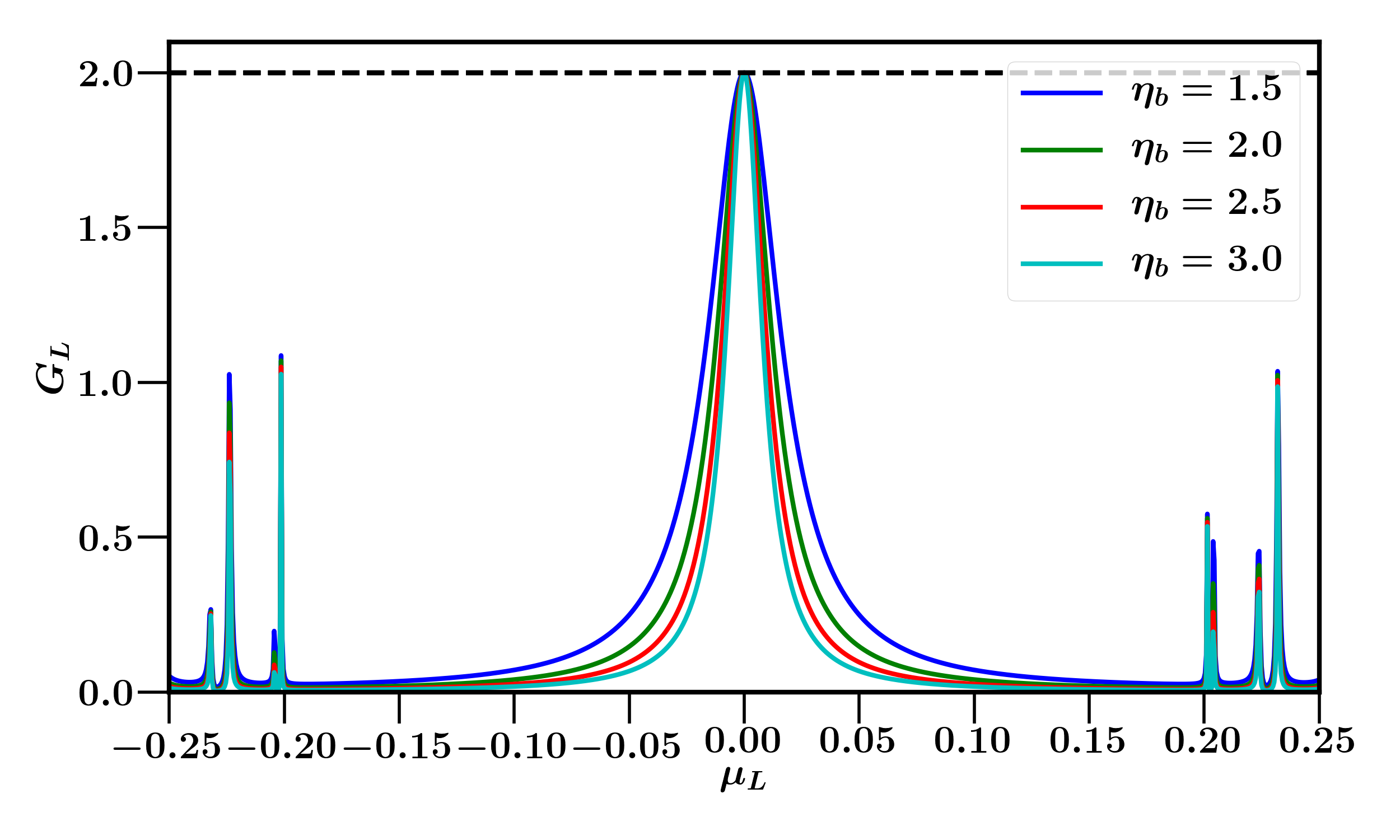}
 	}
 	\subfigure[ Zero bias Peak at different $\mu_w$ with $\eta_b=1.5$, $\eta_c=0.3$ and $\Delta=0.1$]{
 		\includegraphics[width=4.cm,height=33mm]{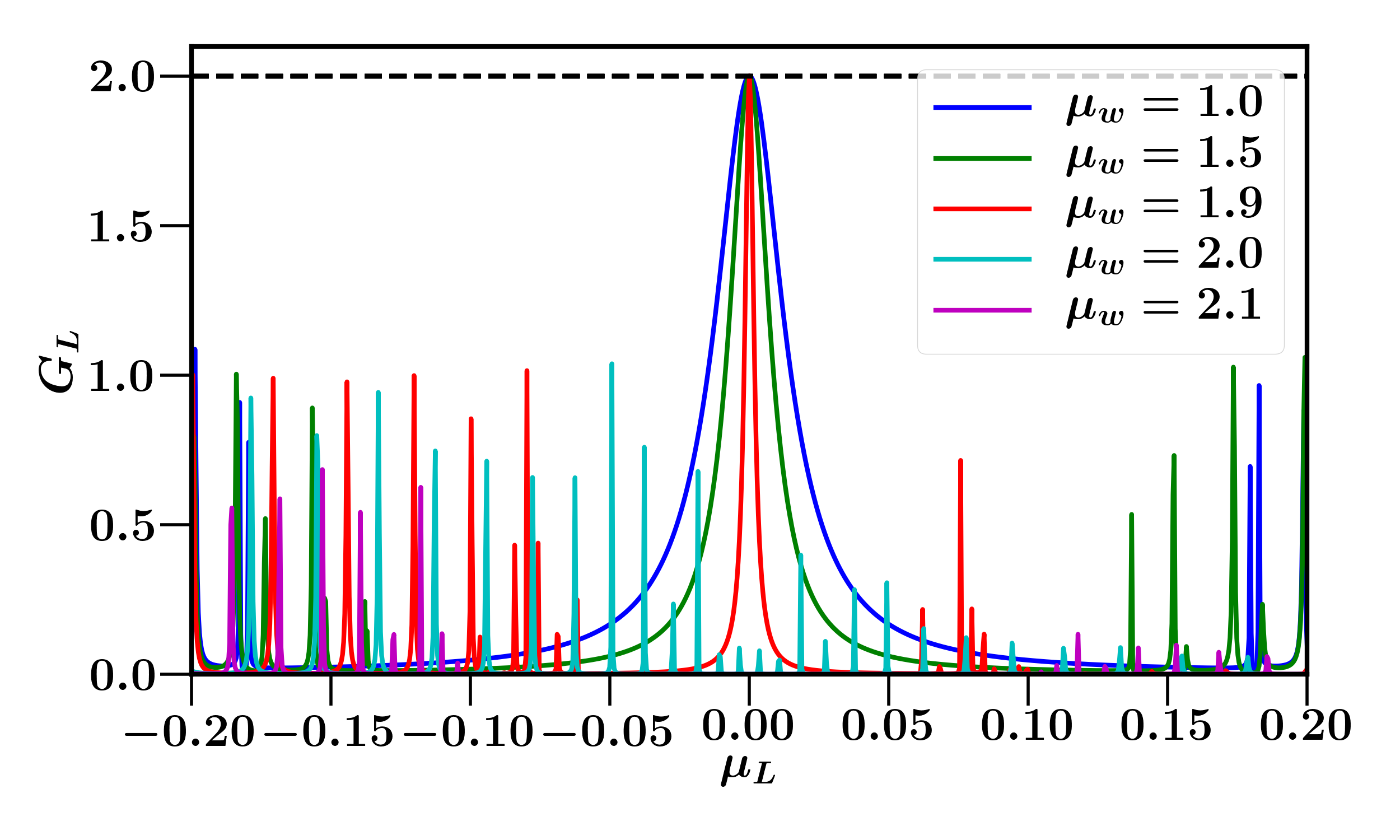}
 	}
 	\subfigure[ Zero bias Peak at different $\Delta$ with $\eta_b=1.5$, $\eta_c=0.3$ and $\mu_w=0.5$]{
 		\includegraphics[width=4.cm,height=33mm]{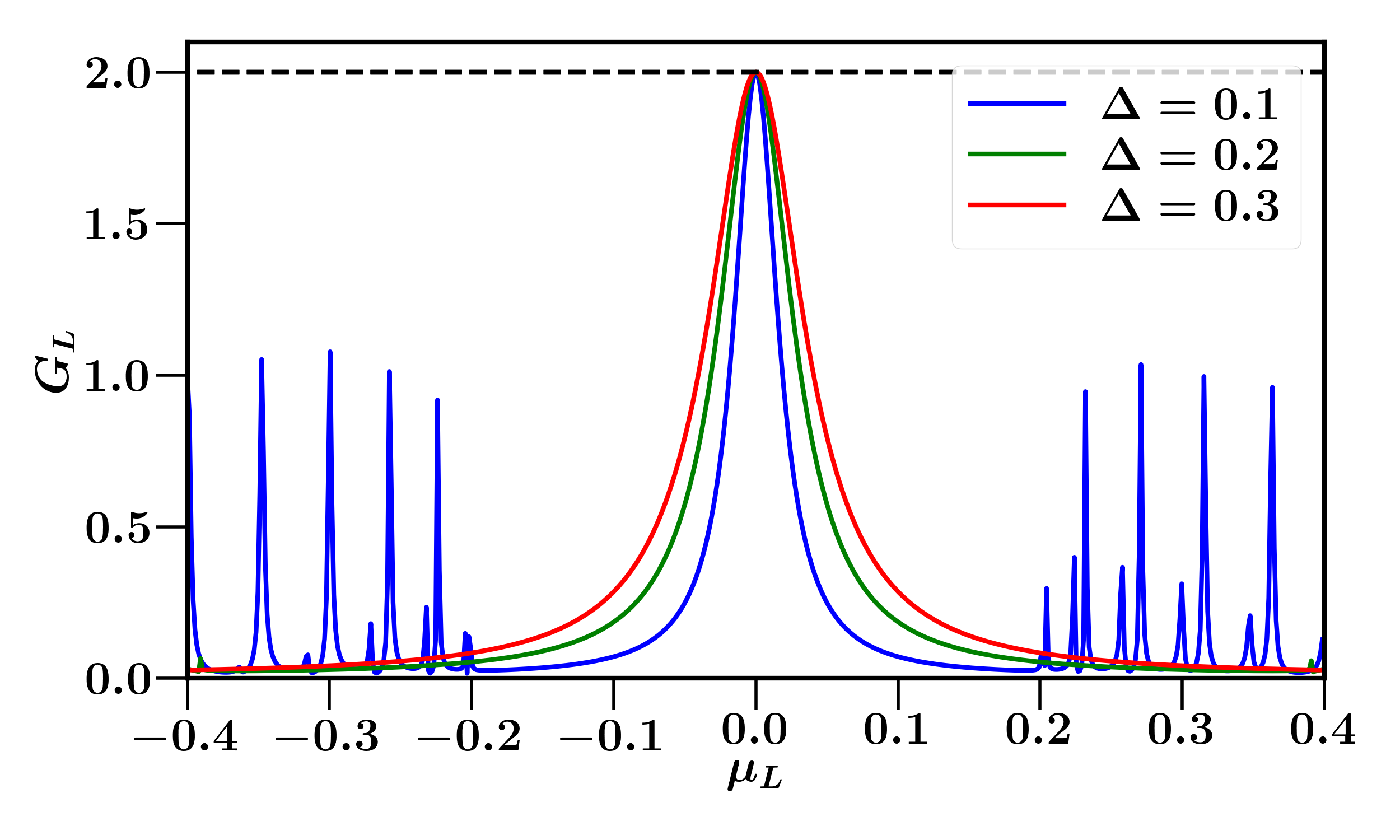}
 	}
 	\caption{Behaviour of the zero bias peak over different parameters of the Hamiltonian. Plots (a) to (d)  are for   $N=100$ and $\eta_w=1$. All other parameters are mentioned in the sub captions of the figures. }
 	\label{zbp_plts}
 \end{figure}

In Fig.~(\ref{specplts}a) and Fig~(\ref{specplts}b) we show respectively the spectrum of the isolated Kitaev wire and the wire connected to reservoirs.  The isolated band spectrum agrees with the expected form $\epsilon_k=\pm \sqrt{(\mu_w+ 2 \eta_w \cos k)^2+4 \Delta^2 \sin^2 k}$, $k \in (-\pi,\pi)$.   In Fig.~(\ref{specplts}c) and Fig.~(\ref{specplts}d), we plot two  wavefunctions of the system, one whose energy lies in the gap of the isolated wire and the other just outside the gap. We see that the first is almost fully localized in the reservoirs while the second is localized in the wire. 
 In Figs.~(\ref{mbsplts}) we plot the zero-energy state in the gap and as expected this gives us the MBS wavefunction, mostly localized at the two edges inside the wire but with some leakage into the leads. The plots in Fig.~(\ref{mbsplts}a) and Fig.~(\ref{mbsplts}b) show a comparison between the numerically obtained MBS and the analytical ones given by Eq.~(\ref{mbspsir}-\ref{mbs1pssiw}) and we find perfect agreement.  The plots in Fig.~(\ref{mbsplts}c) and Fig.~(\ref{mbsplts}d) show the MBS wavefunction for different parameters of the wire Hamiltonian.  We will see shortly that the wavefunctions in Fig.~(\ref{specplts}) and Fig.~(\ref{mbsplts})  help in understanding different features of Fig.~(\ref{plts}) and also Fig.~(\ref{zbp_plts}) (which shows the behaviour of the zero bias conductance peak for different parameters). 
 

The features seen in the spectrum in Figs.~(\ref{specplts},\ref{mbsplts}) can be related to the results for conductance in Fig.~\eqref{plts}. 
  The zero bias peak in the conductance is due to the perfect Andreev reflection$(|r_a|^2=1)$, supported by the Majorana bound states of the wire~\cite{roy2012,mourik2012,maiellaro2019,das2012evidence}. For long wires, these bound states exist in the topological parameter regime, $(\mu_w<2\eta_w)$.  The zero energy MBS in the isolated wire can be clearly seen in its spectrum,  Fig.~(\ref{specplts}a), at the middle of the superconducting gap (SC gap), while for the wire connected to reservoirs, Fig.~(\ref{specplts}b), it is hidden in the contimuum spectrum due to the leads. Plotting the wavefunctions, Figs.~(\ref{mbsplts}a,b), reveals their nature. The plots in Fig.~(\ref{plts}e and \ref{plts}g) tell us $t_n$ and $t_a$ are zero within a certain  energy range which  is  the same as the SC gap for the isolated wire. This makes sense  physically as for long wires the transmission from the left bath to right bath could only happen via the excitation of finite energy quasiparticles   of the superconducting wire which are not possible if the energy of the incoming particle is within the SC gap. Typical wavefunctions in  the superconducting gap look like Fig.~(\ref{specplts}c), having most of their support in the two baths and this explains why there is no transmission from the left reservoir to the right reservoir and we get $t_n=t_a=0$. Outside the SC gap, on the other hand, there are  wavefunctions  which look like Fig.~(\ref{specplts}d) having most of their support inside the wire with some 
leakage into the baths. These thus contribute to the transmission peaks outside the gap.

 In Figs.~(\ref{zbp_plts}a-\ref{zbp_plts}d) we show how the width of the zero bias conductance peak  varies as we vary different parameters of the Hamiltonian. Let us consider Fig.~(\ref{zbp_plts}a) and Fig.~(\ref{zbp_plts}b) which show the zero bias peak for different $\eta_c$ and $\eta_b$ respectively while keeping all other parameters fixed. We see that the peak broadens as we increase $\eta_c$ or decrease $\eta_b$. This can be understood in terms of the MBS wavefunction  given by Eq.~(\ref{mbs1psir}-\ref{mbs1pssiw}) where we saw that the height of the peak in the MBS is proportional to $\frac{\eta_b}{\eta_c}$. Hence, decreasing $\eta_b$ or increasing $\eta_c$ would  decrease  the height of the peak and therefore increase  the weight of the MBS in the left reservoir. 
 Thus the peak broadens as the weight of the MBS in the reservoirs increases. Similarly, from Fig.~(\ref{zbp_plts}c) and Fig.~(\ref{zbp_plts}d) we see that the peak broadens  as $\mu_w$  decreases or $\Delta$  increases and correspondingly the weight of the wavefunction in the reservoirs, Fig.~(\ref{mbsplts}c) and Fig.~(\ref{mbsplts}d), increases. From Fig.~(\ref{zbp_plts}c) we see that the peak splits  at the transition point~\cite{das2012zero}, $(\mu_w=2\eta_w)$, marking the topological phase transition into a topologically trivial state, and then disappears.

\section{Conclusion}
\label{sec_concl}
In conclusion, we provided an analytical proof of  the equivalence of the QLE-NEGF approach and the scattering approach to electron transport in a 1-D superconducting wire. In both cases we start from the same microscopic model of a Kitaev wire connected to one-dimensional leads (baths) containing free Fermions in equilibrium. In the former method one starts with the Heisenberg equations of motion of the full system and eliminates the bath degrees of freedom to obtain effective quantum Langevin equations of motion. The steady state solutuon of these leads to the NEGF formula for the conductance  in terms of a set of nonequilibrium Green's functions.   In the second approach one considers the wire as a scatterer of  plane waves from the leads and writes down the corresponding scattering  solutions for the energy eigenstates. These solutions involve scattering amplitudes  that are obtained using the boundary conditions at the wire-leads junctions. The conductance at the junction is then given by the net probability of transmission of particles across the junction.

We summarize here our some of the main results:
\begin{itemize}
\item We obtained the complete solution of the scattering states in the Kitaev chain, including closed form expressions for the scattering amplitudes $t_a,t_n,r_n,r_a$. 
\item We  obtained the special zero energy solution corresponding to the  MBS state of the isolated open Kitaev chain. We showed that this state exists in the same parameter regime as for the isolated wire. 
\item The conductance of the wire from the QLE-NEGF method and the scattering approach are  given respectively by  Eq.~(\ref{GLnegf1}) and Eq.~(\ref{GLscat1}).   We showed analytically  that the terms in the NEGF conductance expression, $T_1(E)$, $T_2(E)$ and $T_3(E)$, can be related to the  scattering amplitudes $t_n$, $t_a$ and $r_a$ respectively. This leads us to proving the complete equivalence of the two formulas for conductance and hence of the two approaches. 
 
\item We have demonstrated clearly and explicitly the physical interpretation--- from our derivation we see that  the expression for current, Eq.~(\ref{currexp1}), is  exactly in   Landauer's form with each of the baths playing the role of a "double reservoir",  of   electrons and holes. The wire acts as a scatterer and scatters the incoming electrons as holes or electrons into the two baths. Therefore, an electron from say the left bath may end up  being scattered as a hole or an electron into the left bath only. These two processes are the normal reflection and Andreev reflection processes respectively. The  electron may also end up being scattered into the right reservoir, therefore transmitted across the wire, as an electron or a hole. Out of the four possibilities of the future of an electron from the left bath, all excepting the normal reflection  lead to particles being transmitted across the left junction. Therefore, only these three actually contribute to the conductance of the wire. This is the reason behind the NEGF current expression having three distinct terms with the probabilities of these processes  multiplied with the corresponding difference of thermal occupations of the incoming electrons and outgoing electron or holes as one  typically finds in Landauer expressions.     
\item Finally we have given numerical examples that (a) demonstrate the equivalence of the two approaches, (b) show the nature of scattering wavefunctions and the MBS state and their dependence on various parameters and (c) relate the conductance properties to those of the spectrum. 
\end{itemize}

\bibliography{biblio}

%merlin.mbs apsrev4-1.bst 2010-07-25 4.21a (PWD, AO, DPC) hacked
%Control: key (0)
%Control: author (8) initials jnrlst
%Control: editor formatted (1) identically to author
%Control: production of article title (-1) disabled
%Control: page (0) single
%Control: year (1) truncated
%Control: production of eprint (0) enabled
\begin{thebibliography}{20}%
\makeatletter
\providecommand \@ifxundefined [1]{%
 \@ifx{#1\undefined}
}%
\providecommand \@ifnum [1]{%
 \ifnum #1\expandafter \@firstoftwo
 \else \expandafter \@secondoftwo
 \fi
}%
\providecommand \@ifx [1]{%
 \ifx #1\expandafter \@firstoftwo
 \else \expandafter \@secondoftwo
 \fi
}%
\providecommand \natexlab [1]{#1}%
\providecommand \enquote  [1]{``#1''}%
\providecommand \bibnamefont  [1]{#1}%
\providecommand \bibfnamefont [1]{#1}%
\providecommand \citenamefont [1]{#1}%
\providecommand \href@noop [0]{\@secondoftwo}%
\providecommand \href [0]{\begingroup \@sanitize@url \@href}%
\providecommand \@href[1]{\@@startlink{#1}\@@href}%
\providecommand \@@href[1]{\endgroup#1\@@endlink}%
\providecommand \@sanitize@url [0]{\catcode `\\12\catcode `\$12\catcode
  `\&12\catcode `\#12\catcode `\^12\catcode `\_12\catcode `\%12\relax}%
\providecommand \@@startlink[1]{}%
\providecommand \@@endlink[0]{}%
\providecommand \url  [0]{\begingroup\@sanitize@url \@url }%
\providecommand \@url [1]{\endgroup\@href {#1}{\urlprefix }}%
\providecommand \urlprefix  [0]{URL }%
\providecommand \Eprint [0]{\href }%
\providecommand \doibase [0]{http://dx.doi.org/}%
\providecommand \selectlanguage [0]{\@gobble}%
\providecommand \bibinfo  [0]{\@secondoftwo}%
\providecommand \bibfield  [0]{\@secondoftwo}%
\providecommand \translation [1]{[#1]}%
\providecommand \BibitemOpen [0]{}%
\providecommand \bibitemStop [0]{}%
\providecommand \bibitemNoStop [0]{.\EOS\space}%
\providecommand \EOS [0]{\spacefactor3000\relax}%
\providecommand \BibitemShut  [1]{\csname bibitem#1\endcsname}%
\let\auto@bib@innerbib\@empty
%</preamble>
\bibitem [{\citenamefont {Kitaev}(2001)}]{kitaev}%
  \BibitemOpen
  \bibfield  {author} {\bibinfo {author} {\bibfnamefont {A.~Y.}\ \bibnamefont
  {Kitaev}},\ }\href@noop {} {\bibfield  {journal} {\bibinfo  {journal}
  {Physics-Uspekhi}\ }\textbf {\bibinfo {volume} {44}},\ \bibinfo {pages} {131}
  (\bibinfo {year} {2001})}\BibitemShut {NoStop}%
\bibitem [{\citenamefont {Sau}\ \emph {et~al.}(2010)\citenamefont {Sau},
  \citenamefont {Tewari}, \citenamefont {Lutchyn}, \citenamefont {Stanescu},\
  and\ \citenamefont {Sarma}}]{sau2010non}%
  \BibitemOpen
  \bibfield  {author} {\bibinfo {author} {\bibfnamefont {J.~D.}\ \bibnamefont
  {Sau}}, \bibinfo {author} {\bibfnamefont {S.}~\bibnamefont {Tewari}},
  \bibinfo {author} {\bibfnamefont {R.~M.}\ \bibnamefont {Lutchyn}}, \bibinfo
  {author} {\bibfnamefont {T.~D.}\ \bibnamefont {Stanescu}}, \ and\ \bibinfo
  {author} {\bibfnamefont {S.~D.}\ \bibnamefont {Sarma}},\ }\href@noop {}
  {\bibfield  {journal} {\bibinfo  {journal} {Physical Review B}\ }\textbf
  {\bibinfo {volume} {82}},\ \bibinfo {pages} {214509} (\bibinfo {year}
  {2010})}\BibitemShut {NoStop}%
\bibitem [{\citenamefont {Oreg}\ \emph {et~al.}(2010)\citenamefont {Oreg},
  \citenamefont {Refael},\ and\ \citenamefont {von Oppen}}]{oreg2010helical}%
  \BibitemOpen
  \bibfield  {author} {\bibinfo {author} {\bibfnamefont {Y.}~\bibnamefont
  {Oreg}}, \bibinfo {author} {\bibfnamefont {G.}~\bibnamefont {Refael}}, \ and\
  \bibinfo {author} {\bibfnamefont {F.}~\bibnamefont {von Oppen}},\ }\href@noop
  {} {\bibfield  {journal} {\bibinfo  {journal} {Physical review letters}\
  }\textbf {\bibinfo {volume} {105}},\ \bibinfo {pages} {177002} (\bibinfo
  {year} {2010})}\BibitemShut {NoStop}%
\bibitem [{\citenamefont {Mourik}\ \emph {et~al.}(2012)\citenamefont {Mourik},
  \citenamefont {Zuo}, \citenamefont {Frolov}, \citenamefont {Plissard},
  \citenamefont {Bakkers},\ and\ \citenamefont {Kouwenhoven}}]{mourik2012}%
  \BibitemOpen
  \bibfield  {author} {\bibinfo {author} {\bibfnamefont {V.}~\bibnamefont
  {Mourik}}, \bibinfo {author} {\bibfnamefont {K.}~\bibnamefont {Zuo}},
  \bibinfo {author} {\bibfnamefont {S.~M.}\ \bibnamefont {Frolov}}, \bibinfo
  {author} {\bibfnamefont {S.}~\bibnamefont {Plissard}}, \bibinfo {author}
  {\bibfnamefont {E.~P.}\ \bibnamefont {Bakkers}}, \ and\ \bibinfo {author}
  {\bibfnamefont {L.~P.}\ \bibnamefont {Kouwenhoven}},\ }\href@noop {}
  {\bibfield  {journal} {\bibinfo  {journal} {Science}\ }\textbf {\bibinfo
  {volume} {336}},\ \bibinfo {pages} {1003} (\bibinfo {year}
  {2012})}\BibitemShut {NoStop}%
\bibitem [{\citenamefont {Das}\ \emph {et~al.}(2012{\natexlab{a}})\citenamefont
  {Das}, \citenamefont {Ronen}, \citenamefont {Most}, \citenamefont {Oreg},
  \citenamefont {Heiblum},\ and\ \citenamefont {Shtrikman}}]{das2012zero}%
  \BibitemOpen
  \bibfield  {author} {\bibinfo {author} {\bibfnamefont {A.}~\bibnamefont
  {Das}}, \bibinfo {author} {\bibfnamefont {Y.}~\bibnamefont {Ronen}}, \bibinfo
  {author} {\bibfnamefont {Y.}~\bibnamefont {Most}}, \bibinfo {author}
  {\bibfnamefont {Y.}~\bibnamefont {Oreg}}, \bibinfo {author} {\bibfnamefont
  {M.}~\bibnamefont {Heiblum}}, \ and\ \bibinfo {author} {\bibfnamefont
  {H.}~\bibnamefont {Shtrikman}},\ }\href@noop {} {\bibfield  {journal}
  {\bibinfo  {journal} {Nature Physics}\ }\textbf {\bibinfo {volume} {8}},\
  \bibinfo {pages} {887} (\bibinfo {year} {2012}{\natexlab{a}})}\BibitemShut
  {NoStop}%
\bibitem [{\citenamefont {Thakurathi}\ \emph {et~al.}(2015)\citenamefont
  {Thakurathi}, \citenamefont {Deb},\ and\ \citenamefont {Sen}}]{diptiman2015}%
  \BibitemOpen
  \bibfield  {author} {\bibinfo {author} {\bibfnamefont {M.}~\bibnamefont
  {Thakurathi}}, \bibinfo {author} {\bibfnamefont {O.}~\bibnamefont {Deb}}, \
  and\ \bibinfo {author} {\bibfnamefont {D.}~\bibnamefont {Sen}},\ }\href@noop
  {} {\bibfield  {journal} {\bibinfo  {journal} {Journal of Physics: Condensed
  Matter}\ }\textbf {\bibinfo {volume} {27}},\ \bibinfo {pages} {275702}
  (\bibinfo {year} {2015})}\BibitemShut {NoStop}%
\bibitem [{\citenamefont {Roy}\ \emph {et~al.}(2012)\citenamefont {Roy},
  \citenamefont {Bolech},\ and\ \citenamefont {Shah}}]{roy2012}%
  \BibitemOpen
  \bibfield  {author} {\bibinfo {author} {\bibfnamefont {D.}~\bibnamefont
  {Roy}}, \bibinfo {author} {\bibfnamefont {C.}~\bibnamefont {Bolech}}, \ and\
  \bibinfo {author} {\bibfnamefont {N.}~\bibnamefont {Shah}},\ }\href@noop {}
  {\bibfield  {journal} {\bibinfo  {journal} {Physical Review B}\ }\textbf
  {\bibinfo {volume} {86}},\ \bibinfo {pages} {094503} (\bibinfo {year}
  {2012})}\BibitemShut {NoStop}%
\bibitem [{\citenamefont {Bondyopadhaya}\ and\ \citenamefont
  {Roy}(2019)}]{roy2019}%
  \BibitemOpen
  \bibfield  {author} {\bibinfo {author} {\bibfnamefont {N.}~\bibnamefont
  {Bondyopadhaya}}\ and\ \bibinfo {author} {\bibfnamefont {D.}~\bibnamefont
  {Roy}},\ }\href {\doibase 10.1103/PhysRevB.99.214514} {\bibfield  {journal}
  {\bibinfo  {journal} {Phys. Rev. B}\ }\textbf {\bibinfo {volume} {99}},\
  \bibinfo {pages} {214514} (\bibinfo {year} {2019})}\BibitemShut {NoStop}%
\bibitem [{\citenamefont {Bhat}\ and\ \citenamefont
  {Dhar}(2020)}]{bhat2020transport}%
  \BibitemOpen
  \bibfield  {author} {\bibinfo {author} {\bibfnamefont {J.~M.}\ \bibnamefont
  {Bhat}}\ and\ \bibinfo {author} {\bibfnamefont {A.}~\bibnamefont {Dhar}},\
  }\href {\doibase 10.1103/PhysRevB.102.224512} {\bibfield  {journal} {\bibinfo
   {journal} {Phys. Rev. B}\ }\textbf {\bibinfo {volume} {102}},\ \bibinfo
  {pages} {224512} (\bibinfo {year} {2020})}\BibitemShut {NoStop}%
\bibitem [{\citenamefont {Blonder}\ \emph {et~al.}(1982)\citenamefont
  {Blonder}, \citenamefont {Tinkham},\ and\ \citenamefont {Klapwijk}}]{btk}%
  \BibitemOpen
  \bibfield  {author} {\bibinfo {author} {\bibfnamefont {G.~E.}\ \bibnamefont
  {Blonder}}, \bibinfo {author} {\bibfnamefont {M.}~\bibnamefont {Tinkham}}, \
  and\ \bibinfo {author} {\bibfnamefont {T.~M.}\ \bibnamefont {Klapwijk}},\
  }\href {\doibase 10.1103/PhysRevB.25.4515} {\bibfield  {journal} {\bibinfo
  {journal} {Phys. Rev. B}\ }\textbf {\bibinfo {volume} {25}},\ \bibinfo
  {pages} {4515} (\bibinfo {year} {1982})}\BibitemShut {NoStop}%
\bibitem [{\citenamefont {Maiellaro}\ \emph {et~al.}(2019)\citenamefont
  {Maiellaro}, \citenamefont {Romeo}, \citenamefont {Perroni}, \citenamefont
  {Cataudella},\ and\ \citenamefont {Citro}}]{maiellaro2019}%
  \BibitemOpen
  \bibfield  {author} {\bibinfo {author} {\bibfnamefont {A.}~\bibnamefont
  {Maiellaro}}, \bibinfo {author} {\bibfnamefont {F.}~\bibnamefont {Romeo}},
  \bibinfo {author} {\bibfnamefont {C.~A.}\ \bibnamefont {Perroni}}, \bibinfo
  {author} {\bibfnamefont {V.}~\bibnamefont {Cataudella}}, \ and\ \bibinfo
  {author} {\bibfnamefont {R.}~\bibnamefont {Citro}},\ }\href@noop {}
  {\bibfield  {journal} {\bibinfo  {journal} {Nanomaterials}\ }\textbf
  {\bibinfo {volume} {9}},\ \bibinfo {pages} {894} (\bibinfo {year}
  {2019})}\BibitemShut {NoStop}%
\bibitem [{\citenamefont {Lobos}\ and\ \citenamefont
  {Sarma}(2015)}]{lobos2015}%
  \BibitemOpen
  \bibfield  {author} {\bibinfo {author} {\bibfnamefont {A.~M.}\ \bibnamefont
  {Lobos}}\ and\ \bibinfo {author} {\bibfnamefont {S.~D.}\ \bibnamefont
  {Sarma}},\ }\href {\doibase 10.1088/1367-2630/17/6/065010} {\bibfield
  {journal} {\bibinfo  {journal} {New Journal of Physics}\ }\textbf {\bibinfo
  {volume} {17}},\ \bibinfo {pages} {065010} (\bibinfo {year}
  {2015})}\BibitemShut {NoStop}%
\bibitem [{\citenamefont {Doornenbal}\ \emph {et~al.}(2015)\citenamefont
  {Doornenbal}, \citenamefont {Skantzaris},\ and\ \citenamefont
  {Stoof}}]{doornenbal2015conductance}%
  \BibitemOpen
  \bibfield  {author} {\bibinfo {author} {\bibfnamefont {R.}~\bibnamefont
  {Doornenbal}}, \bibinfo {author} {\bibfnamefont {G.}~\bibnamefont
  {Skantzaris}}, \ and\ \bibinfo {author} {\bibfnamefont {H.}~\bibnamefont
  {Stoof}},\ }\href@noop {} {\bibfield  {journal} {\bibinfo  {journal}
  {Physical Review B}\ }\textbf {\bibinfo {volume} {91}},\ \bibinfo {pages}
  {045419} (\bibinfo {year} {2015})}\BibitemShut {NoStop}%
\bibitem [{\citenamefont {Komnik}(2016)}]{komnik2016}%
  \BibitemOpen
  \bibfield  {author} {\bibinfo {author} {\bibfnamefont {A.}~\bibnamefont
  {Komnik}},\ }\href@noop {} {\bibfield  {journal} {\bibinfo  {journal}
  {Physical Review B}\ }\textbf {\bibinfo {volume} {93}},\ \bibinfo {pages}
  {125117} (\bibinfo {year} {2016})}\BibitemShut {NoStop}%
\bibitem [{\citenamefont {Zhang}\ and\ \citenamefont {Quan}(2020)}]{zhang2020}%
  \BibitemOpen
  \bibfield  {author} {\bibinfo {author} {\bibfnamefont {F.}~\bibnamefont
  {Zhang}}\ and\ \bibinfo {author} {\bibfnamefont {H.}~\bibnamefont {Quan}},\
  }\href@noop {} {\bibfield  {journal} {\bibinfo  {journal} {arXiv preprint
  arXiv:2011.05823}\ } (\bibinfo {year} {2020})}\BibitemShut {NoStop}%
\bibitem [{\citenamefont {Dhar}\ and\ \citenamefont {Sen}(2006)}]{dhar2006}%
  \BibitemOpen
  \bibfield  {author} {\bibinfo {author} {\bibfnamefont {A.}~\bibnamefont
  {Dhar}}\ and\ \bibinfo {author} {\bibfnamefont {D.}~\bibnamefont {Sen}},\
  }\href@noop {} {\bibfield  {journal} {\bibinfo  {journal} {Physical Review
  B}\ }\textbf {\bibinfo {volume} {73}},\ \bibinfo {pages} {085119} (\bibinfo
  {year} {2006})}\BibitemShut {NoStop}%
\bibitem [{\citenamefont {Nehra}\ \emph {et~al.}(2020)\citenamefont {Nehra},
  \citenamefont {Sharma},\ and\ \citenamefont {Soori}}]{nehra2020}%
  \BibitemOpen
  \bibfield  {author} {\bibinfo {author} {\bibfnamefont {R.}~\bibnamefont
  {Nehra}}, \bibinfo {author} {\bibfnamefont {A.}~\bibnamefont {Sharma}}, \
  and\ \bibinfo {author} {\bibfnamefont {A.}~\bibnamefont {Soori}},\
  }\href@noop {} {\bibfield  {journal} {\bibinfo  {journal} {EPL (Europhysics
  Letters)}\ }\textbf {\bibinfo {volume} {130}},\ \bibinfo {pages} {27003}
  (\bibinfo {year} {2020})}\BibitemShut {NoStop}%
\bibitem [{\citenamefont {Das}\ and\ \citenamefont
  {Dhar}(2012)}]{das2012landauer}%
  \BibitemOpen
  \bibfield  {author} {\bibinfo {author} {\bibfnamefont {S.~G.}\ \bibnamefont
  {Das}}\ and\ \bibinfo {author} {\bibfnamefont {A.}~\bibnamefont {Dhar}},\
  }\href@noop {} {\bibfield  {journal} {\bibinfo  {journal} {The European
  Physical Journal B}\ }\textbf {\bibinfo {volume} {85}},\ \bibinfo {pages}
  {372} (\bibinfo {year} {2012})}\BibitemShut {NoStop}%
\bibitem [{\citenamefont {Das}\ \emph {et~al.}(2012{\natexlab{b}})\citenamefont
  {Das}, \citenamefont {Ronen}, \citenamefont {Most}, \citenamefont {Oreg},
  \citenamefont {Heiblum},\ and\ \citenamefont {Shtrikman}}]{das2012evidence}%
  \BibitemOpen
  \bibfield  {author} {\bibinfo {author} {\bibfnamefont {A.}~\bibnamefont
  {Das}}, \bibinfo {author} {\bibfnamefont {Y.}~\bibnamefont {Ronen}}, \bibinfo
  {author} {\bibfnamefont {Y.}~\bibnamefont {Most}}, \bibinfo {author}
  {\bibfnamefont {Y.}~\bibnamefont {Oreg}}, \bibinfo {author} {\bibfnamefont
  {M.}~\bibnamefont {Heiblum}}, \ and\ \bibinfo {author} {\bibfnamefont
  {H.}~\bibnamefont {Shtrikman}},\ }\href@noop {} {\bibfield  {journal}
  {\bibinfo  {journal} {arXiv preprint arXiv:1205.7073}\ } (\bibinfo {year}
  {2012}{\natexlab{b}})}\BibitemShut {NoStop}%
\bibitem [{\citenamefont {Aguado}(2017)}]{aguado2017majorana}%
  \BibitemOpen
  \bibfield  {author} {\bibinfo {author} {\bibfnamefont {R.}~\bibnamefont
  {Aguado}},\ }\href@noop {} {\bibfield  {journal} {\bibinfo  {journal} {La
  Rivista del Nuovo Cimento}\ }\textbf {\bibinfo {volume} {40}},\ \bibinfo
  {pages} {523} (\bibinfo {year} {2017})}\BibitemShut {NoStop}%
\end{thebibliography}%
\end{document}